\documentclass[journal]{IEEEtran}

\usepackage{cite}      

\usepackage{amsmath}
\usepackage{amssymb}
\usepackage{amsfonts}
\usepackage{graphicx}
\usepackage{array}
\usepackage{color}
\usepackage{bbold}
\usepackage[font=footnotesize]{caption}
\usepackage{subcaption}
\usepackage{enumitem}
\usepackage{float}
\usepackage{url}
\usepackage[normalem]{ulem}
\setlist[enumerate]{label*=\arabic*.}
\newcommand{\be}{\begin{equation}}
\newcommand{\ee}{\end{equation}}
\newcommand{\bea}{\begin{eqnarray}}
\newcommand{\eea}{\end{eqnarray}}
\newcommand{\beaa}{\begin{eqnarray*}}
\newcommand{\eeaa}{\end{eqnarray*}}
\newcommand{\ben}{\begin{enumerate}}
\newcommand{\een}{\end{enumerate}}
\newcommand{\bi}{\begin{itemize}}
\newcommand{\ei}{\end{itemize}}

\newcommand{\uu}{\underline}

\newcommand{\eps}{\varepsilon}

\newcommand{\df}{{\rm d}}

\begin{document}

\title{Iterative Symbol Recovery For Power Efficient DC Biased Optical OFDM Systems}

\author{Amir Weiss, Arie Yeredor, and Mark Shtaif
\thanks{A. Weiss, A. Yeredor and M. Shtaif are with the School of Electrical Engineering, Faculty of Engineering, Tel Aviv University, Tel Aviv 69978, Israel (e-mail: amirweiss15@gmail.com; arie@eng.tau.ac.il; shtaif@eng.tau.ac.il).}}


\maketitle

\begin{abstract}
Orthogonal frequency division multiplexing (OFDM) has proven itself as an effective multi-carrier digital communication technique. In recent years the interest in optical OFDM
has grown significantly, due to its spectral efficiency and inherent resilience to frequency-selective channels and to narrowband interference. {For these reasons it is
currently considered to be one of the leading candidates for deployment in short fiber links such as the ones intended for inter-data-center communications.} In this paper we
present a new power-efficient symbol recovery scheme for dc-biased optical OFDM (DCO-OFDM) in an intensity-modulation direct-detection (IM/DD) system. We introduce an
alternative method for clipping in order to maintain a non-negative
real-valued signal and still preserve information, which is lost when using clipping, and propose an iterative detection algorithm for this method. A reduction of $50\%$ in the transmitted optical power along with an increase of signal-independent noise immunity (gaining $3$[dB] in SNR), compared to traditional DCO-OFDM with a DC bias of $2$ standard deviations of the OFDM signal, is attained by our new scheme for a symbol error rate (SER) of $10^{-3}$ in a QPSK constellation additive white gaussian noise (AWGN) flat channel model.\\
\end{abstract}

\begin{IEEEkeywords}
Optical Communication, DC-Biased Optical Orthogonal Frequency Division Multiplexing (DCO-OFDM), Power Efficient.
\end{IEEEkeywords}

\section{Introduction}
Although orthogonal frequency division multiplexing (OFDM) has been known for its advantages in quite a while \cite{chang1970orthogonal,salz1969fourier,peled1980frequency}, only
in recent years it has turned into a major field of interest in optical communications \cite{dixon2001orthogonal,armstrong2009ofdm}. {Particularly attractive is the
intensity-modulation direct-detection (IM/DD) scheme, which, owing to its simple implementation and low cost, constitutes one of the leading candidates for inter-data-center
communications}. For such systems, the data is modulated onto an electrical signal, $x(t)$, which can be either voltage or current, depending on the details of the electrical
system, and an optical intensity modulator generates an optical signal with intensity $\alpha x(t)$, where $\alpha$ is a real-valued positive coefficient. We emphasize that the
light's intensity, and not its amplitude, is proportional to the electrical signal. Therefore, $x(t)$ must be a real-valued non-negative signal $x(t)\in \Re_{\text{\tiny{$+$}}} \
\forall t$, which generally does not hold for an unconstrained baseband OFDM signal. In order to generate a real-valued baseband OFDM signal, the vector of symbols of length $N$
(derived from a given constellation scheme), representing the bits, must meet the constraint of Hermitian symmetry, thereby sacrificing half of the allocated bandwidth. After
generating the real-valued signal, further modification is needed so as to meet the non-negativity condition. To this end, two primary methods have been proposed: asymmetrically
clipped OFDM (ACO-OFDM) \cite{armstrong2006power,armstrong2008comparison} and dc-biased optical OFDM (DCO-OFDM) \cite{carruthers1996multiple,gonzalez2005ofdm}. ACO-OFDM is a
method suggested by Armstrong and Lowery \cite{armstrong2006power} where the OFDM signal is generated such that only the odd frequency subcarriers carry the information (i.e.
symbols from a given constellation), and the even frequency subcarriers are set to zero. After the IFFT is applied, the signal is clipped at the zero level. In this form, all of
the clipping noise falls on the even subcarriers, and the data carrying odd subcarriers remain undistorted. This method has been shown to be efficient in terms of optical power
and from an information theoretic perspective \cite{li2007channel}, however it uses only one quarter of the available bandwidth, hence it is wasteful in terms of spectral
efficiency. DCO-OFDM is a method in which a DC bias is added to the real-valued baseband OFDM signal in order to reduce the probability that it assumes negative values. Since,
owing to the high peak to average power ratio (PAPR) characterizing OFDM, the positivity of the signal cannot be guaranteed with $100\%$ certainty, the signal is clipped (i.e. its
value is replaced by zero) in the intervals in which it is negative. The clipping operation causes distortion (mostly referred to as clipping noise) which limits the performance
(in terms of symbol error rate (SER))\cite{armstrong2008comparison}. The DC bias is set so as to ensure that the probability of a clipping event, and consequently the impact of
clipping noise, remain acceptably low. The disadvantage of the DCO-OFDM is that the DC bias significantly increases the transmitted optical power. Several methods were proposed
with the aim of addressing the power efficiency problem and coping with the implications of clipping noise. Most notable are the unipolar OFDM (U-OFDM) \cite{tsonev2012novel},
which uses different time sample states and a rearrangement of the OFDM frame; asymmetrically companded DCO-OFDM (ADO-OFDM) \cite{barrami2014optical}, which uses a linear
companding function in order to compress the negative part of the signal, thereby reducing the clipping noise, the Hartley Transform is proposed in \cite{svaluto2010novel} for
optical IM/DD OFDM, and more. In this paper, we propose a new approach for meeting the non-negativity condition, which is based on replacing the negative parts of the DC-biased
OFDM signal with their absolute value. In this case the traditional clipping noise is replaced with a stronger noise contribution, but one that is correlated with the signal. By
applying an iterative signs estimation algorithm (ISEA) which will be presented as well, our approach enables a significant reduction in the DC bias level, and a corresponding
improvement in power efficiency. In section II we review the mathematical model of an IM/DD channel and describe the proposed alternative to the clipping method. In section III we
show simulation results where we compare the new suggested method to traditional DCO-OFDM and to a theoretical lower bound on performance. Section IV is devoted to conclusions.

\section{A New Approach For Meeting The Non-Negativity Constraint}
We denote by $s(t)$ the data-carrying intensity waveform that is produced by the transmitter. After being photo-detected and electronically amplified, the electrical signal that
impinges upon the analog to digital conversion unit (ADC) is given by
\begin{equation}\label{eq5}
y(t)=s(t)+v(t),
\end{equation}
where $v(t)$ is zero mean additive white Gaussian noise (AWGN) with variance $\sigma_v^2$. Equation \eqref{eq5} contains a simplified description of the optical channel, as it
assumes that optical propagation effects, most prominently loss and chromatic dispersion, are compensated for prior to photo-detection. In addition, noise following from possible
optical amplification is assumed to be negligible. It is assumed that the ADC resolution is sufficiently high in order to justify the neglect of quantization noise. Finally, Since
a host of standard synchronization methods for optical OFDM communication systems is available \cite{tang2003synchronization,jin2011optical}, we shall follow common practice by
assuming perfect synchronization.
\subsection{An Alternative to Clipping}
Let $\{\alpha_i\}_{i=0}^{K-1}$ be a discrete, closed set of symbols from a given constellation (e.g., 16-QAM) denoted by $\mathcal{A}$, where $K$ is the constellation size, and
let $\{\theta_k\}_{k=0}^{N-1}$ be the series of symbols to be sent by the OFDM signal, where $\theta_k$ is the symbol corresponding to the $k$'th sub-carrier and $N$ is the FFT
size, such that $N=2^m, m\in\mathbb{N}$. In order for the time-domain signal to be real-valued, the vector of symbols to be sent, $\boldsymbol{\theta}^T=[\theta_0 \ \theta_1 \ ...
\ \theta_{N-1}]$, is constrained to Hermitian symmetry such that,
\begin{equation}\label{eq6}
\theta_k=\theta_{N-k}^{*} \ , \ 0<k<\frac{N}{2}
\end{equation}
and the two components $\theta_0$ and $\theta_{N/2}$ are set to zero, i.e. $\theta_0=\theta_{N/2}=0$. We assume $\{\theta_k\}_{k=0}^{N-1}$ are independent identically distributed
(i.i.d) random variables (RVs), distributed uniformly over the set $\mathcal{A}$, and $E[\theta_k]=0$. Let $\{s[n]\}_{n=0}^{N-1}$ be the discrete-time signal at the IFFT output,
corresponding to the vector of symbols $\boldsymbol{\theta}$, defined as
\begin{align}\label{eq7}
s[n]&=\frac{1}{\sqrt{N}}\sum_{k=0}^{N-1}\theta_k e^{j\frac{2\pi}{N}kn} \\
&=\frac{1}{\sqrt{N}}\sum_{k=1}^{N/2-1}\text{Re}\left\{\theta_k\right\} \cos \left(\frac{2\pi}{N}kn\right) \nonumber \\
&-\frac{1}{\sqrt{N}}\sum_{k=1}^{N/2-1}\text{Im}\left\{\theta_k\right\} \sin \left(\frac{2\pi}{N}kn\right) \ , \ 0\le n\le N-1
\end{align}
where the equality is due to (\ref{eq6}), such that all $s[n]$ are real-valued. By the central limit theorem, if $N$ is large enough, each sample $s[n]$ is approximately Gaussian
distributed, with zero mean and variance $\sigma_s^2$ (derived from the constellation parameters). Once the digital OFDM signal is generated, the analog baseband OFDM signal is
produced by a digital to analog converter (DAC). Assuming ideal reconstruction, by Shannon's interpolation formula we have
\begin{equation}\label{eq8}
s(t)=\sum_{n=-\infty}^\infty s[n]p(t-nT_s)
\end{equation}
where $T_s$ denotes the sampling interval and $p(t)\triangleq {\rm sinc}(t/T_s)$ is the ideal reconstruction kernel. Clearly, the condition $s(t)\in\Re$ is satisfied. At this
point, in order to feed the electrical signal into the optical modulator, it is necessary to satisfy the non-negativity condition as well. In DCO-OFDM, a positive DC bias is
introduced such that
\begin{equation}\label{eq9}
s_B(t)\triangleq s(t)+B_{DC} \ , \ B_{DC}>0
\end{equation}
where $s_B(t)$ is the biased baseband OFDM signal, and $B_{DC}$ is the DC bias constant, commonly prescribed in terms of a proportionality constant $\kappa$, such that
\begin{equation}\label{eq10}
B_{DC}\triangleq \kappa\sigma_s=\kappa\sqrt{E[s^2(t)]} \ , \ \kappa>0
\end{equation}
is proportional to the standard deviation of $s(t)$. We define the bias-index as $\beta\triangleq10\log_{10}(1+\kappa^2)$[dB], where the $\log$'s argument is the ratio between the
power of the biased signal $s_B(t)$ to the power of the original signal $s(t)$, which indicates the increase in power of $s_B(t)$ relative to $s(t)$. Since OFDM signals suffer
from high PAPR, the biased signal still does not necessarily meet the non-negativity condition, which requires further manipulation of $s_B(t)$. A commonly used operation is
clipping (at zero level), defined as
\[ h(x) = \left\{ \begin{array}{ll}
x & \mbox{$x \geq 0$}\\
0 & \mbox{$x < 0$}\end{array} \right. .\]
Applied to the biased signal, the clipping operation yields the clipped signal
\begin{equation}\label{eq11}
s_{c,B}(t) = \left\{ \begin{array}{ll}
s_B(t) & \mbox{$s_B(t) \geq 0$}\\
0 & \mbox{$s_B(t) < 0$}\end{array} \right.
\end{equation}
and $s_{c,B}(t)$ is of course non-negative for all $t$, as required. However, this operation results in an additional noise component, usually referred to as clipping noise, since
the clipped signal can be also described as
\begin{equation}\label{eq12}
s_{c,B}(t) = s_B(t)+n_c(t),
\end{equation}
where
\begin{equation}\label{eq13}
n_c(t)\triangleq \left\{ \begin{array}{ll}
0 & \mbox{$s_B(t) \geq 0$}\\
-s_B(t) & \mbox{$s_B(t) < 0$}\end{array} \right.
\end{equation}
is the clipping noise. The bias and the clipping noise are related to one another: on the one hand, increasing the bias is equivalent to reducing the clipping noise power, but the
cost is of course higher transmitted optical power. On the other hand, reducing the bias will save transmitted optical power, but at the same time will increase the clipping noise
power, which may have an unacceptable effect on performance. The effect of clipping noise on performance can be seen in Armstrong and Schmidt's paper
\cite{armstrong2008comparison} for different bias and constellations. In practice, to achieve standard SER values for large constellations, a high signal to noise ratio (SNR) is
required, so the clipping noise must be very low, therefore $B_{DC}$ must be very large. Usually a bias of at least twice the standard deviation of the signal $s(t)$ must be used
\cite{armstrong2006spc07,svaluto2010power}, in order to reduce the clipping noise effect to tolerable levels (in terms of SER).

Clipping indeed ensures a non-negative signal, but at the expense of losing significant information. We propose to substitute the clipping operator with an absolute value
operator,
\begin{equation}\label{eq15}
s_{a,B}(t) = \left|s_B(t)\right|.
\end{equation}
The signal $s_{a,B}(t)$ still meets the non-negativity constraint, and at the same time it carries information that was lost by the clipping operation, and is not available in the
clipped signal $s_{c,B}(t)$. Of course, this additional information is gained at the cost of losing the sign information of all values for all $t$. Nevertheless, as we demonstrate
in what follows, the use of $s_{a,B}(t)$ turns out to be highly beneficial in the context of the OFDM detection process.
\begin{figure}[]
\centering
    \includegraphics[width=0.5\textwidth]{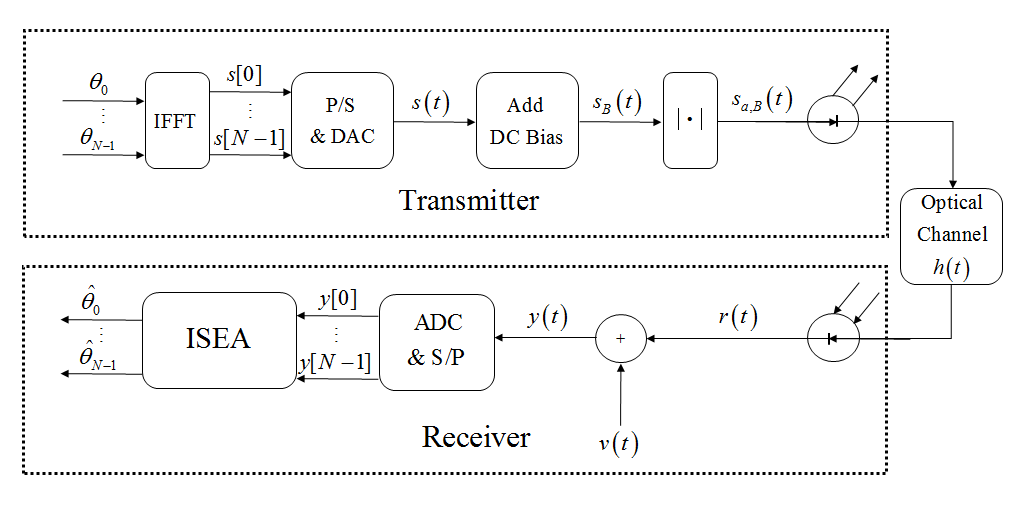}
    \caption{A block diagram of the transmitter and the receiver of the proposed modified DCO-OFDM.}
    \label{fig:Block_Diagram}
\end{figure}
\subsection{System Model}
The block diagram of the proposed modified DCO-OFDM system is presented in Fig. \ref{fig:Block_Diagram}. On the transmitter side, after applying the $N$-size IFFT to the
constellation symbols so as to obtain the discrete-time signal, passing through the DAC and adding a DC bias, the absolute value operation replaces clipping. {Similarly to most
analyses of clipped OFDM systems} \cite{armstrong2008comparison},\cite{barrami2014optical},\cite{armstrong2006spc07}, {we do not include the effects of hardware induced bandwidth
limitations, leaving this item for future study}. After the absolute value is applied, the electrical (real-valued non-negative) baseband signal modulates the optical intensity of
the laser output and the signal is sent through the optical channel. On the receiver side, the measured signal $y(t)$ is sampled by the ADC at the same rate that was used at the
transmitter's DAC. Note that this sampling rate is not the Nyquist rate of $s_{a,B}(t)$, because of the bandwidth-expanding absolute value operation (similarly to the clipping
operation), but this is immaterial since we are not interested in reconstructing $s_{a,B}(t)$. As stated earlier in the beginning of this section, synchronization is assumed.
Hence, we now have
\begin{equation}\label{eq16}
y[n] = s_{a,B}[n]+v[n] \ , \ 0\le n\le N-1
\end{equation}
The samples $\{y[n]\}_{n=0}^{N-1}$ are sent as an input to the iterative signs estimation algorithm (ISEA) block for the symbols recovery process.

\subsection{Iterative Signs Estimation Algorithm}
We start by introducing the ISEA algorithm and its principle of operation while ignoring the presence of additive noise, i.e. with $\sigma_v^2=0$. The performance of the algorithm
in the presence of additive noise will be considered numerically in Sec. \ref{sec:simulresults}.

We will now show how the signs of the samples $\{s_B[n]\}_{n=0}^{N-1}$ are extracted (with high probability) from $\{y[n]\}_{n=0}^{N-1}$, which contains only their absolute
values. Let $\{z[n]\}_{n=0}^{N-1}$ be the (unknown) signs series, corresponding to the series $\{s_B[n]\}_{n=0}^{N-1}$, i.e.
\begin{equation}\label{eq19}
z[n]\triangleq \textrm{sgn}\left(s_B[n]\right)=\left\{ \begin{array}{ll}
+1 & \mbox{$s_B[n] \geq 0$}\\
-1 & \mbox{$s_B[n] < 0$}\end{array} \right.  \ , \ 0\le n\le N-1
\end{equation}
where $0$ is arbitrarily referred to as positive. The series of pairs $\left\{(y[n],z[n])\right\}_{n=0}^{N-1}$ contains complete information about $\{s_B[n]\}_{n=0}^{N-1}$ since
$s_B[n]=z[n]\cdot y[n]$. In order to simplify the exposition, we define the following $N$-dimensional vectors
\begin{equation}\label{eq22}
\begin{split}
&\mathbf{s}=\left[s[0]\;\;s[1]\;...\;s[N-1]\right]^T, \mathbf{z}=\left[z[0]\;\;z[1]\;...\;z[N-1]\right]^T, \\
&\mathbf{y}=\left[y[0]\;\;y[1]\;...\;y[N-1]\right]^T, \boldsymbol{\theta}=\left[\theta_0\;\;\theta_1\;...\;\theta_{N-1}\right]^T.
\end{split}
\end{equation}
The ISEA strives to discover the true values of the series $\{z[n]\}_{n=0}^{N-1}$, and operates as follows:
\begin{enumerate}
  \item (Initialization) set $i \gets 0$ and $\hat{\mathbf{z}}^{(0)}=\mathbb{1}$
  \item Compute the estimated OFDM signal
  \begin{equation}\label{eq23}
  \hat{s}^{(i)}[n]=\hat{z}^{(i)}[n]\cdot y[n] - B_{DC} \ , \  0\le n\le N-1
  \end{equation}
  \item Compute the ``soft" symbols estimators
  \begin{equation}\label{eq24}
  \hat{\tilde{\boldsymbol{\theta}}}^{(i)}=\textrm{FFT}\left\{\hat{\mathbf{s}}^{(i)}\right\}
  \end{equation}
  \item Compute the ``hard" symbols estimators
  \begin{equation}\label{eq25}
  \hat{\boldsymbol{\theta}}^{(i)}=\textrm{Slicer}\left\{\hat{\tilde{\boldsymbol{\theta}}}^{(i)}\right\}
  \end{equation}
  \item If $\hat{\boldsymbol{\theta}}^{(i)}=\hat{\boldsymbol{\theta}}^{(i-1)}$ (and $i\neq0$)
  \begin{enumerate}
  \item return $\hat{\boldsymbol{\theta}}^{(i)}$
  \end{enumerate}
  \item Otherwise,
  \begin{enumerate}
  \item $i \gets i+1$
  \item Compute the updated estimated OFDM signal
  \begin{equation}\label{eq26}
  \hat{\mathbf{s}}^{(i)}=\textrm{IFFT}\left\{\hat{\boldsymbol{\theta}}^{(i-1)}\right\}
  \end{equation}
  \item Compute the updated sign estimators
  \begin{equation}\label{eq27}
  \hat{\mathbf{z}}^{(i)}=\textrm{sgn}\left\{\hat{\mathbf{s}}^{(i)}+\mathbb{1}\cdot B_{DC}\right\}
  \end{equation}
  \item Return to step (2)
  \end{enumerate}
\end{enumerate}
where $\mathbb{1}$ denotes an $N\times 1$ all-ones vector, $\textrm{FFT}\{\cdot\}$ and $\textrm{IFFT}\{\cdot\}$ denote the FFT and IFFF operations, respectively,
$\textrm{Slicer}\{\cdot\}$ denotes the operation of assigning the valid symbol (from the given constellation) with the minimum Euclidian distance to its corresponding ``soft"
estimator, i.e.
\begin{equation}\label{eq28}
\hat{\theta}_k=\underset{\theta \in \mathcal{A}}{\operatorname{argmin}}\left\|\theta-\hat{\tilde{\theta}}_k\right\|^2
\end{equation}
for each entry of $\hat{\boldsymbol{\theta}}$, $\textrm{sgn}\{\cdot\}$ denotes the $\textrm{sgn}(\cdot)$ operation elementwise, and $^{(i)}$ denotes the $i$'th iteration. Note
that generally the ``soft" estimator may be an invalid symbol, i.e. $\hat{\tilde{\theta}}_k\notin \mathcal{A}$.

Analysis of the algorithm in terms of theoretical runtime boundaries and probability of convergence (to the correct solution or at all) is cumbersome, as the complexity of the
problem grows rapidly with the FFT size $N$. For example, an analytical assessment of the probability of an ``Error Event" (upon convergence) defined as
\begin{equation}\label{eq29}
\text{Pr}(\epsilon) \triangleq \text{Pr}\left(\hat{\boldsymbol{\theta}}^{(i)}=\hat{\boldsymbol{\theta}}^{(i-1)}\cap\hat{\boldsymbol{\theta}}^{(i)}\neq\boldsymbol{\theta}\right),
\end{equation}
is highly impractical. In order to do this, one must go over all the possible combinations and find all cases for which the ``soft" estimators of two consecutive iterations,
determined by two (possibly non-identical) sign series, result in the same ``hard" estimator after the Slicer operation. However,  the empirical determination of some useful
parameters regarding the algorithm's performance, for a fixed set of system parameters (e.g., constellation type and size, signal power, DC bias etc.), is both simple and
informative, and as we demonstrate numerically in Sec. \ref{sec:simulresults}, the algorithm converges in only a few iterations to the correct solution for a moderate DC bias.
Intuitively, the algorithm's logic can be explained in the following manner: the absolute values information contained in the measurements reduces the problem to finding the
correct signs. If a sufficient number of samples are assigned with their true original sign, the distortion caused by the absolute value operation, which will be referred to as
absolute value noise (to be further discussed in what follows), will be tolerable (very much like the clipping effect), in the sense that the algorithm will have sufficient
knowledge for correcting the wrongly guessed signs (in the initialization) by the discrete Fourier transform relation between the symbols and the OFDM signal and the prior
information about the valid symbols, derived from the chosen constellation. Notice that in step (2), $\hat{\mathbf{s}}^{(i)}$ is updated by $\hat{\mathbf{z}}^{(i)}$, which is also
updated in each iteration, and by the measurements $\mathbf{y}$, which remain unchanged. In this way, the set of signs being assigned to the measurements is updated iteratively,
until the ``best fit" is found, where ``best fit" is defined by the stopping criterion --- a series of signs for which the algorithm detects in the current iteration a vector of
symbols identical to the detected vector of symbols in the preceding iteration. Obviously, if the initial signs series is the true signs series, the algorithm stops after two
iterations, since no update will occur. The proposed algorithm is somewhat similar in its logic to the DAR algorithm, proposed by Kim and St\"uber in \cite{kim1999clipping}, but
its uniqueness clearly comes from handling the absolute value samples and updating the signs in each iteration by that information. In the next subsection we shall discuss the
distortion caused by the absolute value operation, and shed light on the influence of the key parameters on the proposed method.
\subsection{Absolute Value Noise}
Much like the clipped signal, the absolute value signal can be written as the sum of the biased signal and a noise component, which will be called absolute value (AV) noise so
that
\begin{equation}\label{eq30}
s_{a,B}(t) = s_B(t)+n_a(t),
\end{equation}
where $n_a(t)$ denotes the AV noise, and is defined as
\begin{equation}\label{eq31}
n_a(t)\triangleq \left\{ \begin{array}{ll}
0 & \mbox{$s_B(t) \geq 0$}\\
-2s_B(t) & \mbox{$s_B(t) < 0$}\end{array} \right..
\end{equation}
\begin{figure}[]
\centering
    \includegraphics[width=0.3\textwidth]{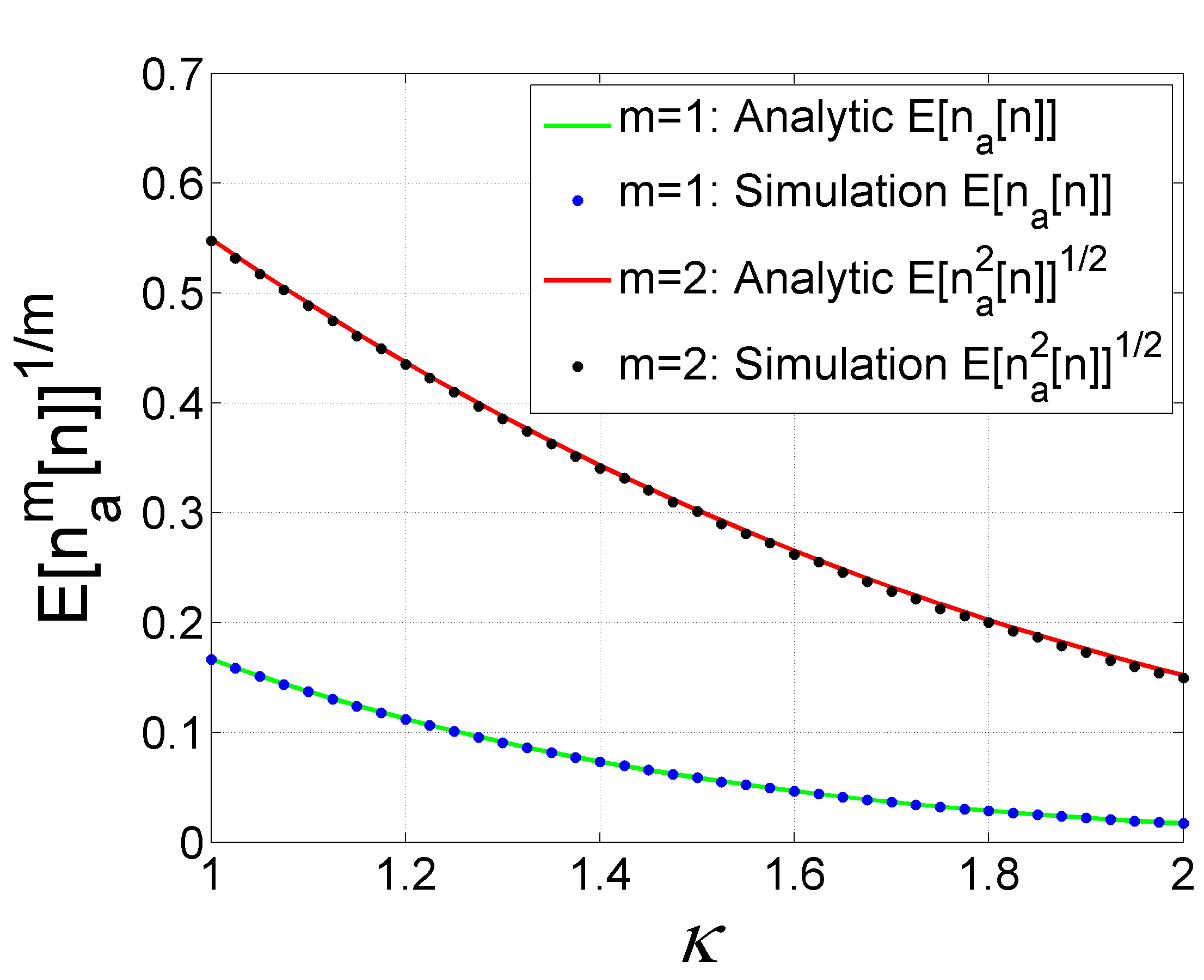}
    \caption{Analytical and simulation results for the AV noise mean and root mean square (RMS) with $N=1024$, $\sigma_s^2=1$ as the system parameters. Simulation results are based on averaging 2000 OFDM realizations, corresponding to a total of $\sim10^6$ symbols.}
    \label{fig:AV_noise_mean_and_rms}
\end{figure}
Note that the AV noise equals exactly two times the noise due to clipping. In a synchronized signal-independent-noiseless system, we have
\begin{equation}\label{eq32}
y[n] = s_B[n]+n_a[n]
\end{equation}
where $n_a[n]$ are the samples of $n_a(t)$. Since by virtue of the central limit theorem, the sample $s[n]$ can be approximated as Gaussian distributed with zero mean and variance
$\sigma_s^2$ (for each $n$), it follows that
\begin{equation}\label{eq34}
s_B[n] \sim N(B_{DC},\sigma_s^2).
\end{equation}
Therefore,
\begin{equation}\label{eq35}
\text{Pr}(s_B[n]<0)= Q\left(\frac{B_{DC}}{\sigma_s}\right)=Q(\kappa)\triangleq p_a,
\end{equation}
where $Q(\cdot)$ denotes the standard Q-function, defined as
\begin{equation}\label{eq36}
Q(x)=\frac{1}{\sqrt{2\pi}}\int\limits_{t}^{\infty}\exp\left(\frac{-t^2}{2}\right)\mathrm{d}t,
\end{equation}
so the AV noise can be characterized by
\begin{equation}\label{eq37}
n_a[n]= \left\{ \begin{array}{ll}
0 & \mbox{$w.p. \ 1-p_a$}\\
-2s_B[n] & \mbox{$w.p. \ \ \ p_a$}\end{array} \right. .
\end{equation}
The mean and power of the AV noise can be analytically computed and are given (in terms of the proportionality constant $\kappa$) by
\begin{equation}\label{eq38}
E\left[n_a[n]\right]=2\sigma_s\left[\frac{1}{\sqrt{2\pi}}e^{\frac{-\kappa^2}{2}}-\kappa\cdot Q(\kappa)\right],
\end{equation}
\begin{equation}\label{eq39}
E\left[n^2_a[n]\right]=4\sigma_s^2\left[(1+\kappa^2)Q(\kappa)-\frac{\kappa}{\sqrt{2\pi}}e^{\frac{-\kappa^2}{2}}\right].
\end{equation}
Fig. \ref{fig:AV_noise_mean_and_rms} shows simulation results that verify these analytical terms, for a true OFDM signal. By this representation, one can learn about the influence
of the DC bias, the original OFDM signal's power and the ratio between them. Notice that when $B_{DC}\gg\sigma_s$, namely $\kappa\gg1$, the AV noise is highly unlikely to appear,
so that $p_a\approx0$. As we demonstrate in Sec. \ref{sec:simulresults}, simulations of the proposed method in a broad range of system parameters reveal a threshold phenomenon
--- for a fixed set of system parameters the algorithm converges to the correct solution if the number of errors in the initial signs assignment does not exceed a certain threshold,
typically much smaller than the FFT size, $N$. Otherwise, the AV noise effect becomes too dominant and the algorithm fails by converging to a false solution. We emphasize that the
threshold number of errors depends on the constellation type and size, and more precisely on the symbol density (i.e. the minimal Euclidian distance between two symbols), and the
FFT size, $N$, because it affects the accuracy of the Gaussianity assumption, on which the procedure relies. In what follows, we denote by $p_{th}$ the ratio between the threshold
number of errors and the FFT size.

The statistical independence between the transmitted data symbols implies that the samples $s_B[n]$ are pairwise uncorrelated. In addition, by virtue of the central limit theorem,
they are approximately Gaussian distributed for large $N$, and hence it is a reasonable approximation to treat them as if they were also statistically independent. In view of the
above, our approach will be to choose the system parameters such that
\begin{equation}\label{eq42}
p_a<p_{th},
\end{equation}
with the idea that this will lead to the convergence of the ISEA algorithm to the correct solution with high probability. Once $p_{th}$ is known, an estimate of the smallest
sufficient bias level can be obtained after substituting Eq. \eqref{eq35} in Eq. \eqref{eq42}, yielding
\begin{equation}\label{eq43}
B_{DC}>\sigma_s\cdot Q^{-1}(p_{th}).
\end{equation}
One can choose the OFDM signal power $\sigma_s^2$ based on the level of the additive noise (so as to guarantee adequate signal to noise ratio) and then use Eq. \eqref{eq43} to
find a bias level that would ensure convergence of the ISEA algorithm. These ideas are further investigated in the following section.

\section{Simulation Results}
\label{sec:simulresults}
\begin{figure}[]
        \centering
        \begin{subfigure}[b]{0.25\textwidth}
                \includegraphics[width=\textwidth]{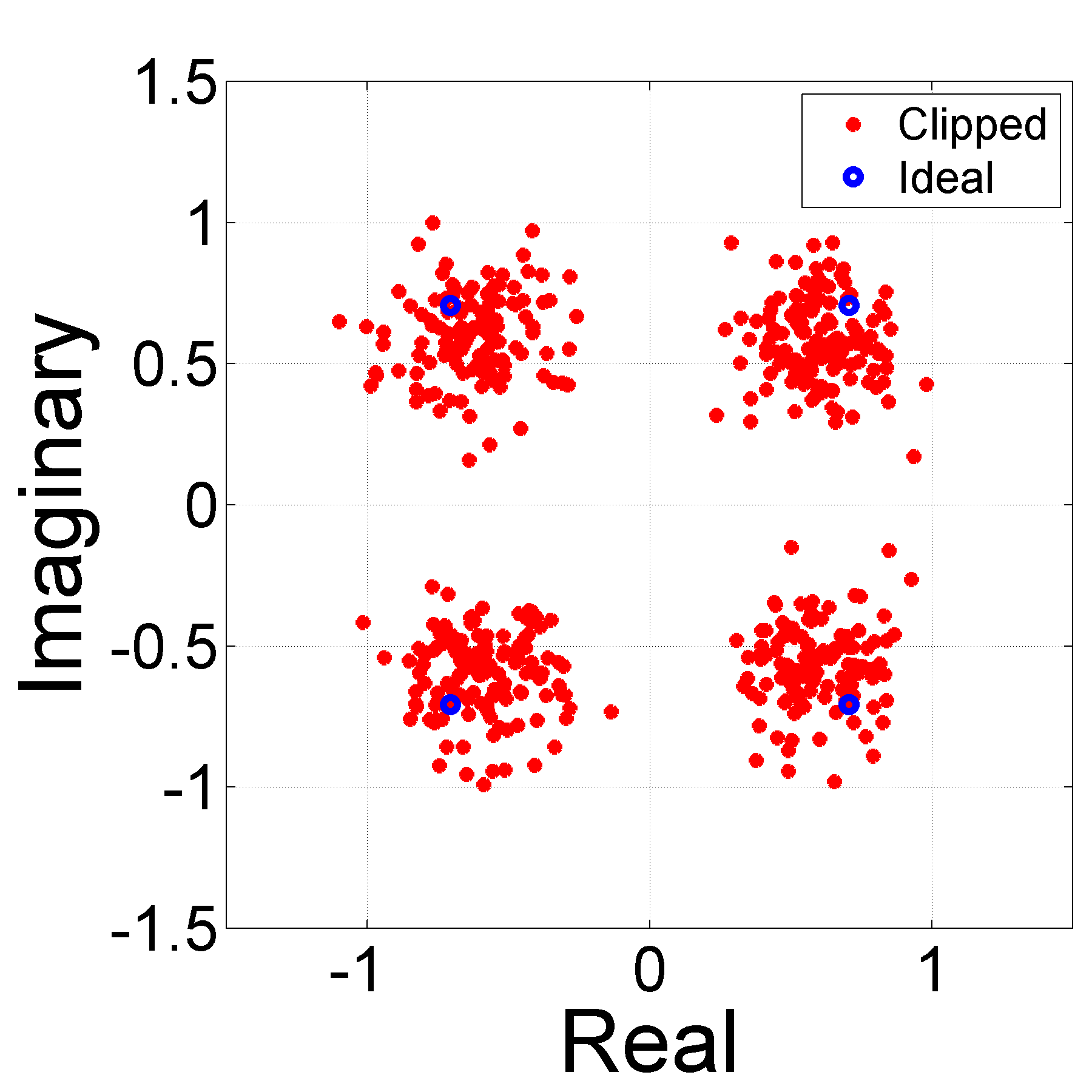}
                \caption{}
                \label{fig:clipping_QPSK}
        \end{subfigure}%
        ~
        \begin{subfigure}[b]{0.25\textwidth}
                \includegraphics[width=\textwidth]{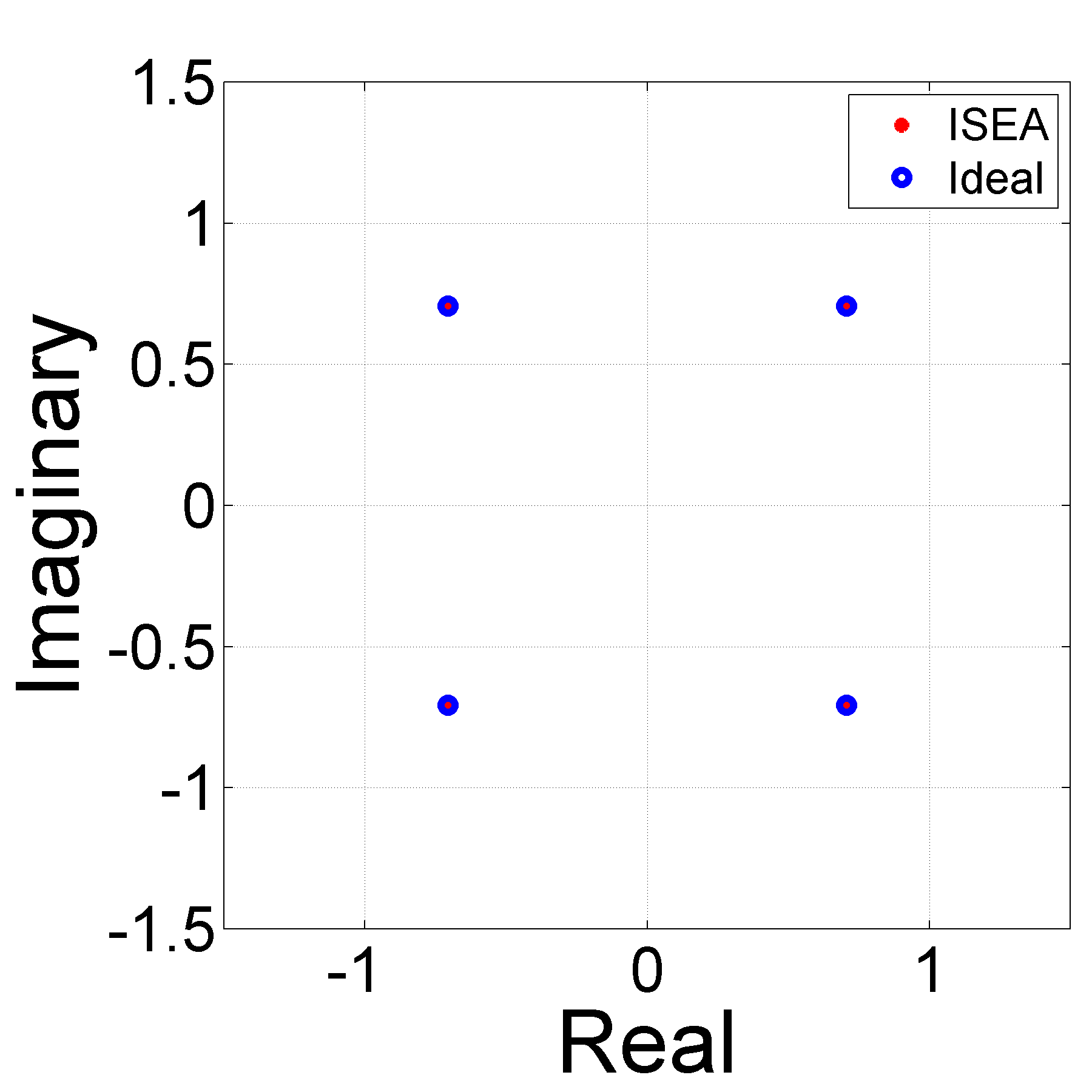}
                \caption{}
                \label{fig:ISEA_QPSK}
        \end{subfigure}
        \caption{Constellation diagram of QPSK. (a) Estimators before slicer for Traditional DCO-OFDM (via clipping) with $\beta=3$[dB] (b) Estimators before slicer for the proposed method (via ISEA) with $\beta=3$[dB]}
        \label{fig:constellationQPSK}
\end{figure}
\begin{figure}[]
        \centering
        \begin{subfigure}[b]{0.25\textwidth}
                \includegraphics[width=\textwidth]{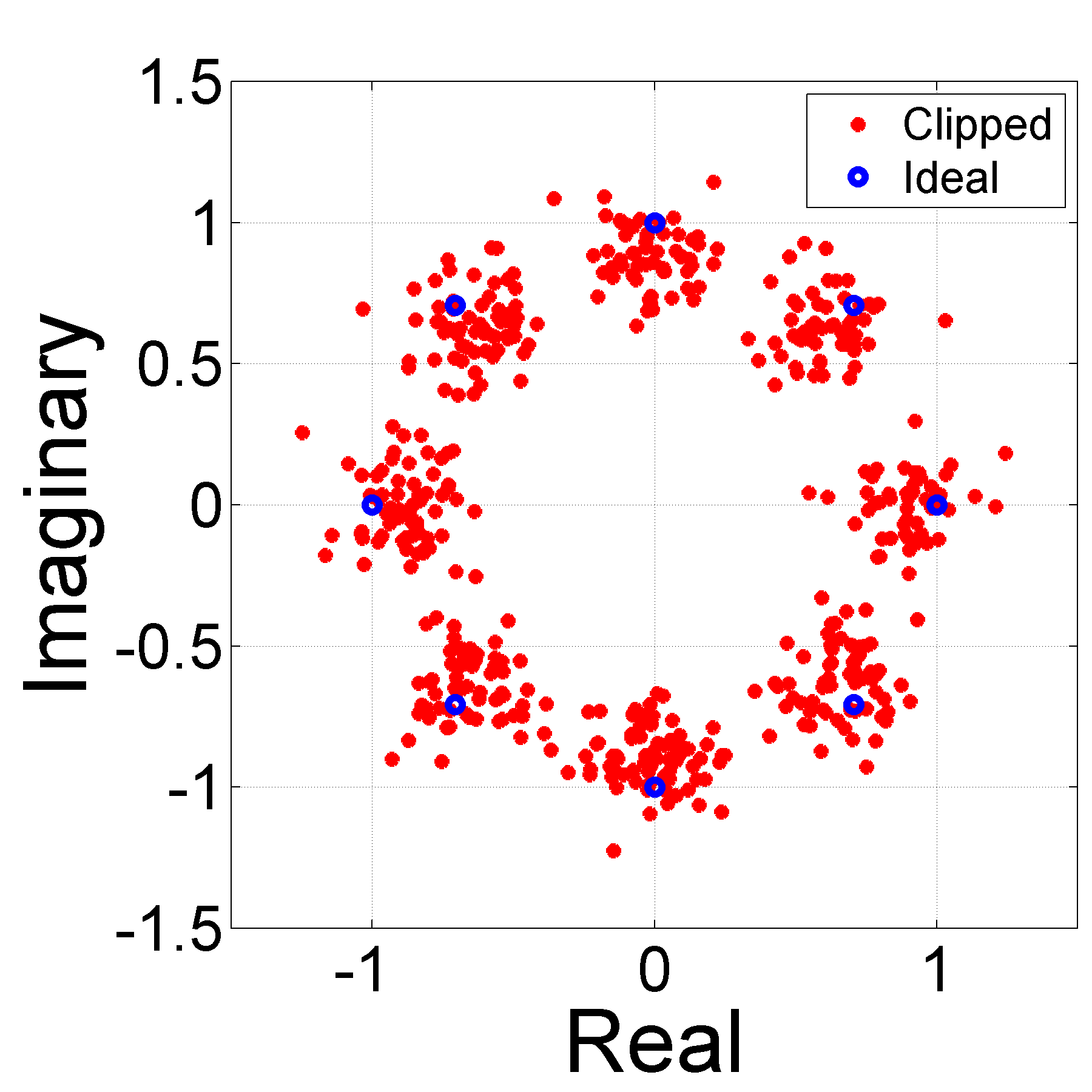}
                \caption{}
                \label{fig:clipping_8PSK}
        \end{subfigure}%
        ~
        \begin{subfigure}[b]{0.25\textwidth}
                \includegraphics[width=\textwidth]{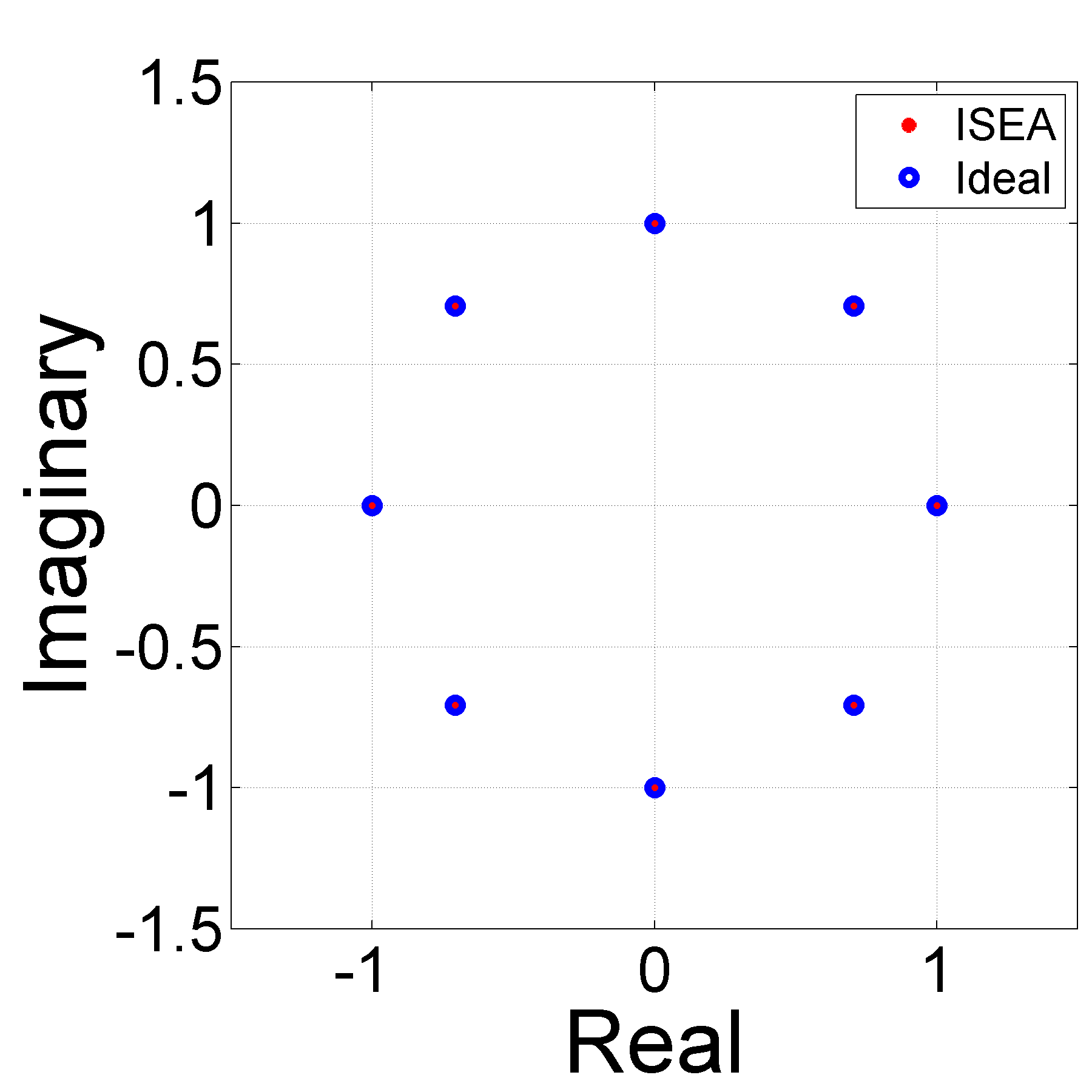}
                \caption{}
                \label{fig:ISEA_8PSK}
        \end{subfigure}
        \caption{Constellation diagram of 8-PSK. (a) Estimators before slicer for Traditional DCO-OFDM (via clipping) with $\beta=4.3$[dB] (b) Estimators before slicer for the proposed method (via ISEA) with $\beta=4.3$[dB]}
        \label{fig:constellation8PSK}
\end{figure}

\begin{figure}[]
        \centering
        \begin{subfigure}[b]{0.25\textwidth}
            \includegraphics[width=\textwidth]{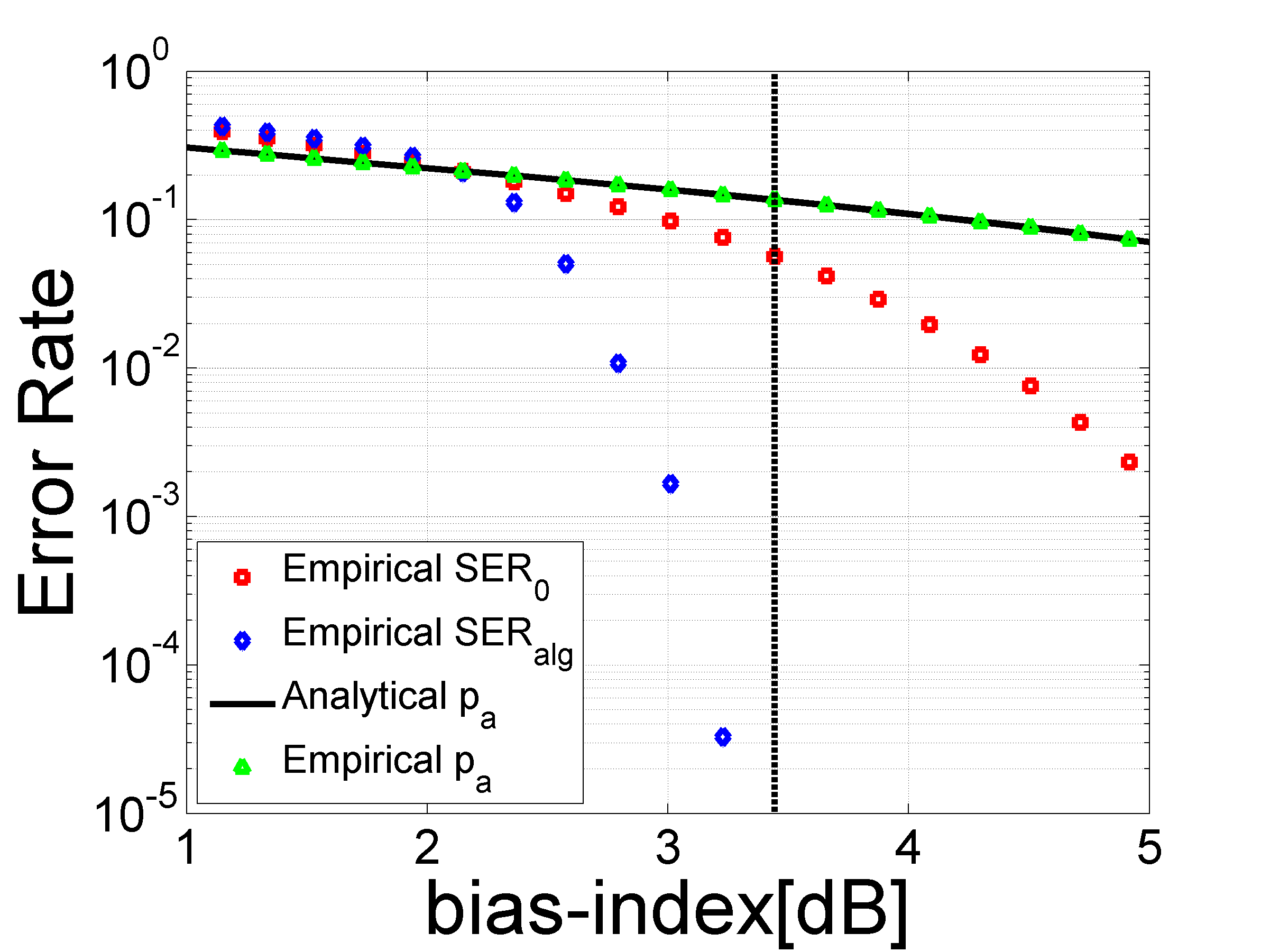}
            \caption{}
            \label{fig:Error_rate_SER0_SER_p_a_QPSK}
        \end{subfigure}%
        ~
        \begin{subfigure}[b]{0.25\textwidth}
            \includegraphics[width=\textwidth]{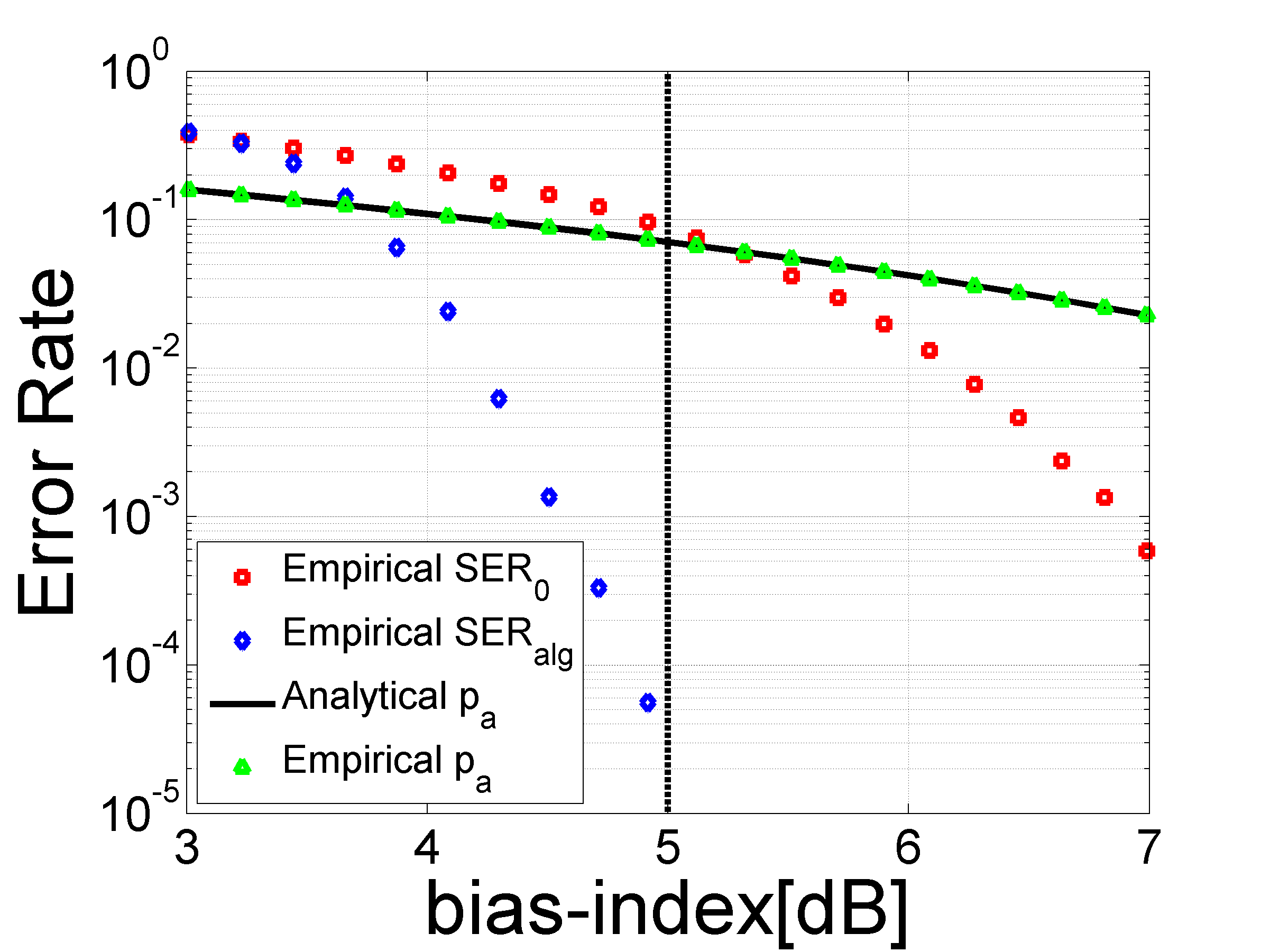}
            \caption{}
            \label{fig:Error_rate_SER0_SER_p_a_8PSK}
        \end{subfigure}
        \caption{Simulation results of initial SER (SER$_0$), ISEA's SER (SER$_{alg}$), analytical and empirical $p_a$ with $N=1024$, $\sigma_s^2=1$ and $\sigma_v^2=0$ based on averaging 4000 OFDM realizations, corresponding to a total of $\sim10^6$ symbols. The horizontal axis is the bias-index defined as $\beta\triangleq10\log_{10}(1+\kappa^2)$. The dashed line marks the bias level above which SER$_{alg}=0$. (a) QPSK constellation (b) $8$-PSK constellation}
        \label{fig:Error_rate_SER0_SER_p_a}
\end{figure}

\begin{figure}[]
        \centering
        \begin{subfigure}[b]{0.25\textwidth}
            \includegraphics[width=\textwidth]{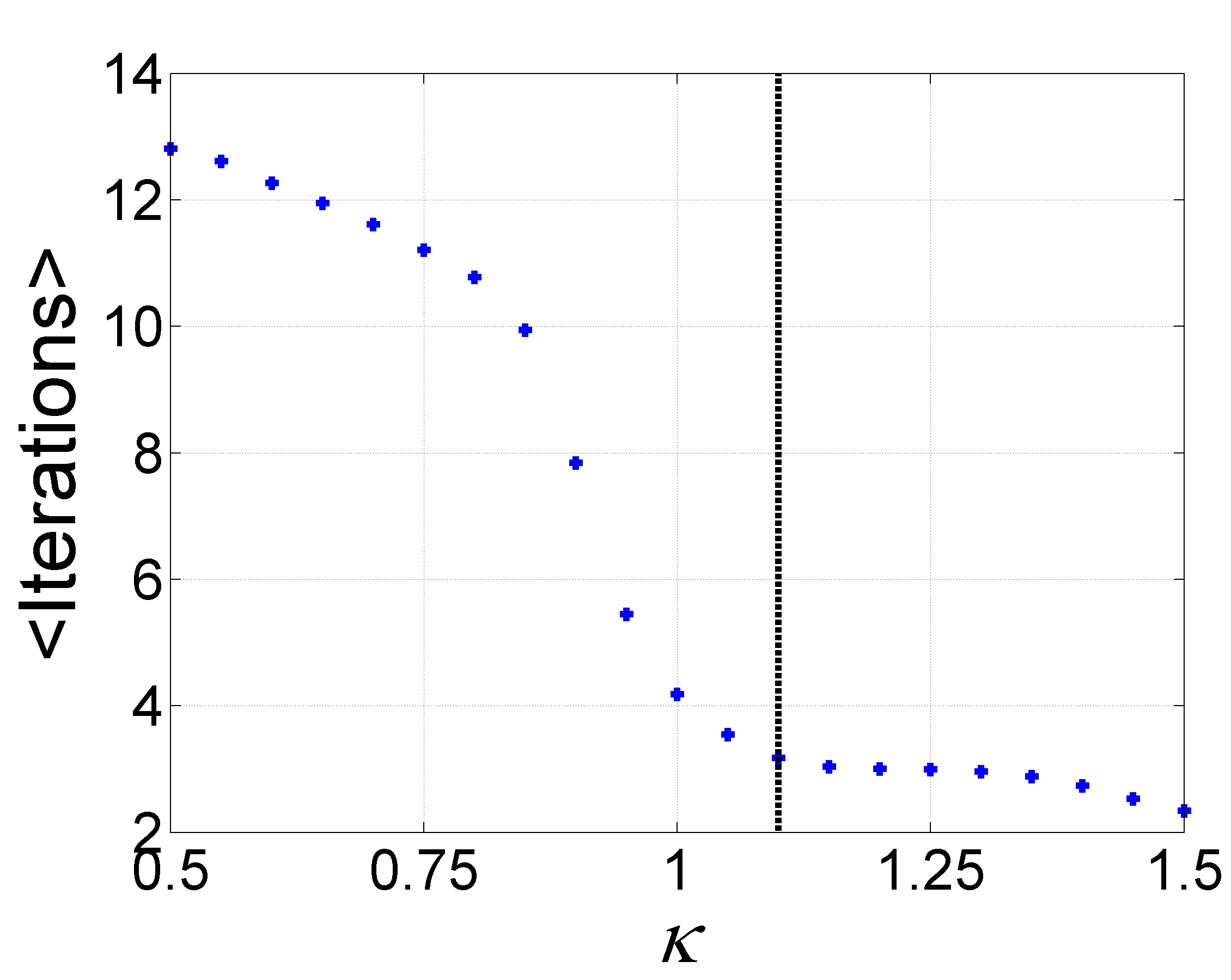}
            \caption{}
            \label{fig:avg_num_of_iter_HQ_QPSK}
        \end{subfigure}%
        ~
        \begin{subfigure}[b]{0.25\textwidth}
           \includegraphics[width=\textwidth]{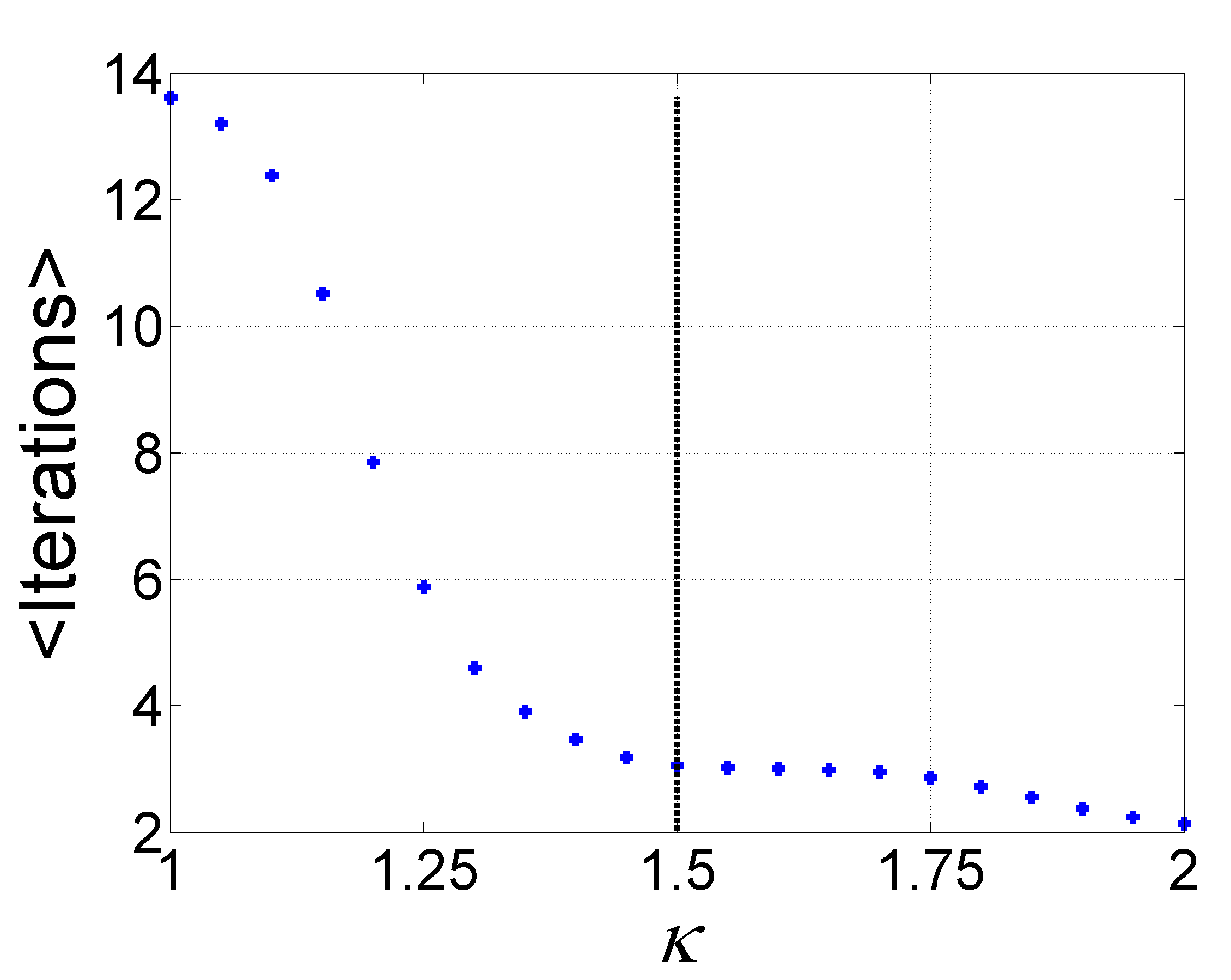}
            \caption{}
            \label{fig:avg_num_of_iter_HQ_8PSK}
        \end{subfigure}
        \caption{The average number of iterations to convergence with $N=1024$, $\sigma_s^2=1$ and $\sigma_v^2=0$ based on averaging 4000 OFDM realizations, corresponding to a total of $\sim10^6$ symbols. The dashed line marks the bias level above which SER$_{alg}=0$. (a) QPSK constellation. (b) $8$-PSK constellation. {The average number of iterations to convergence is 3.0024 with QPSK, and 3.0166 for 8-PSK, obtained with $\kappa=1.2$ and $\kappa=1.55$, respectively. The respective standard deviations were 0.06 and 0.13. In all 4000 realizations the maximum observed number of iterations to convergence was 4.}}
        \label{fig:avg_num_of_iter}
\end{figure}

\begin{figure}[]
        \centering
        \begin{subfigure}[b]{0.25\textwidth}
            \includegraphics[width=\textwidth]{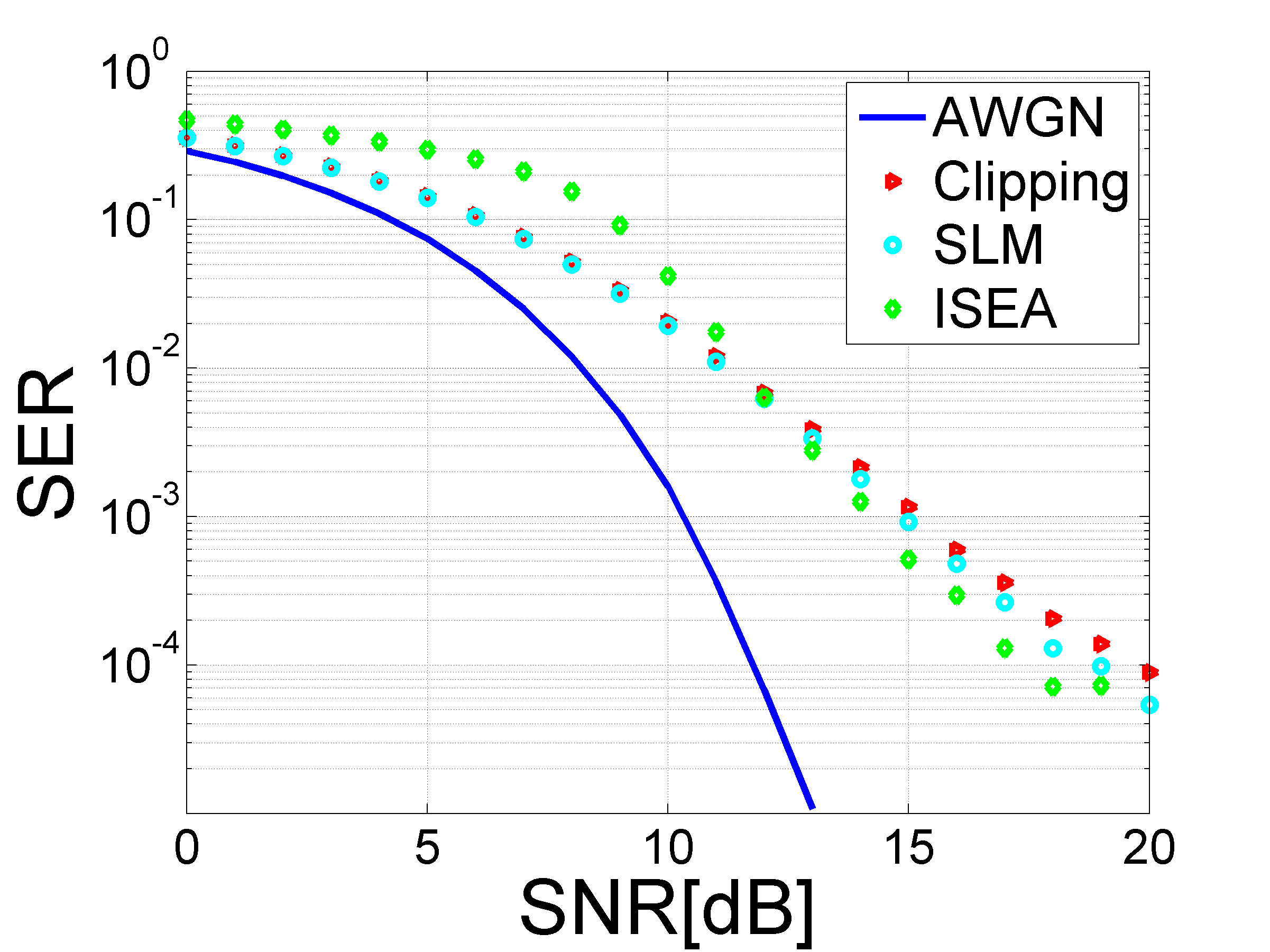}
            \caption{}
            \label{fig:SER_Vs_SNR_k_1_08_HQ_QPSK}
        \end{subfigure}%
        ~
        \begin{subfigure}[b]{0.25\textwidth}
            \includegraphics[width=\textwidth]{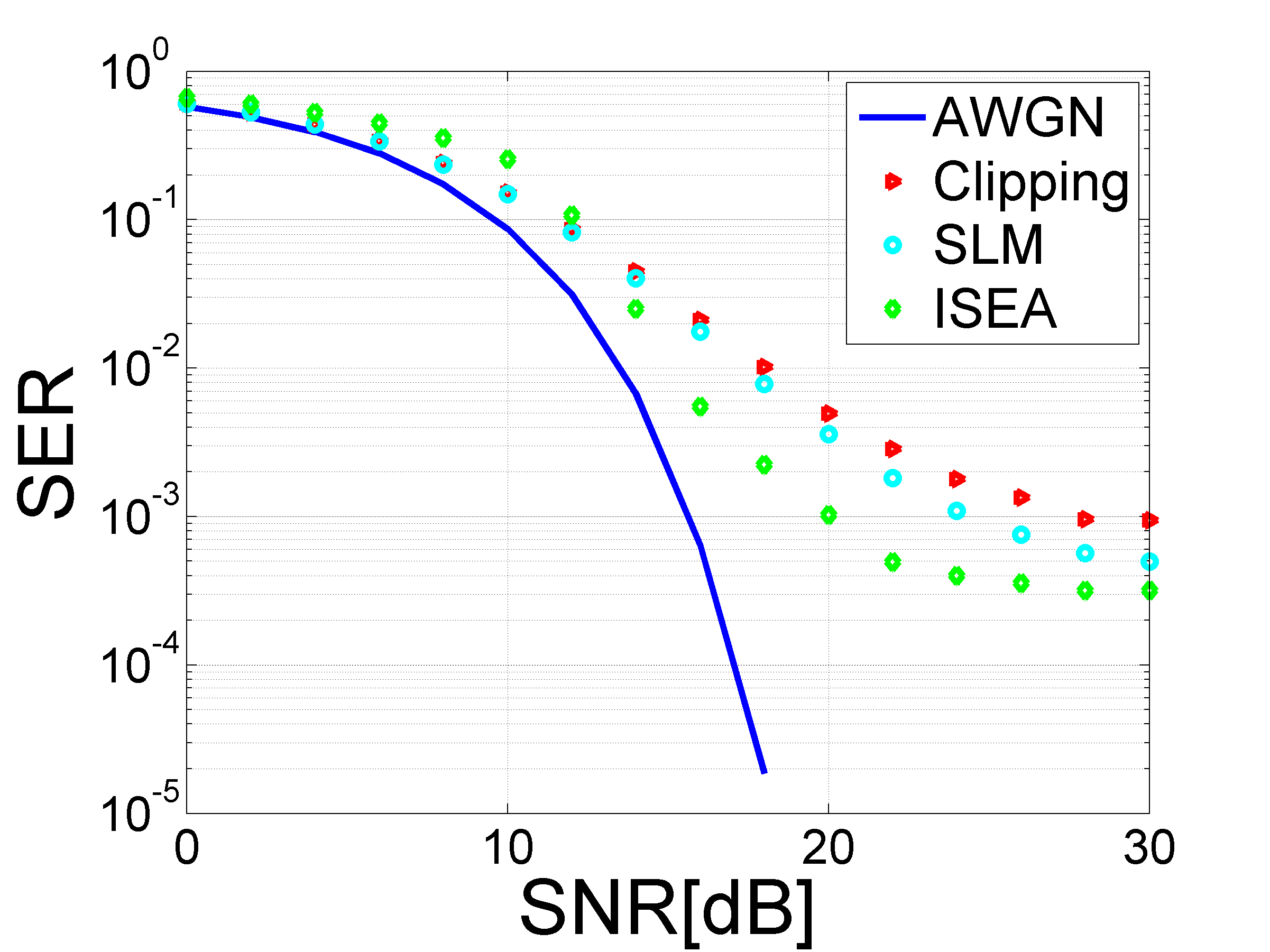}
            \caption{}
            \label{fig:SER_Vs_SNR_k_1_4_HQ_8PSK}
        \end{subfigure}
        \begin{subfigure}[b]{0.25\textwidth}
            \includegraphics[width=\textwidth]{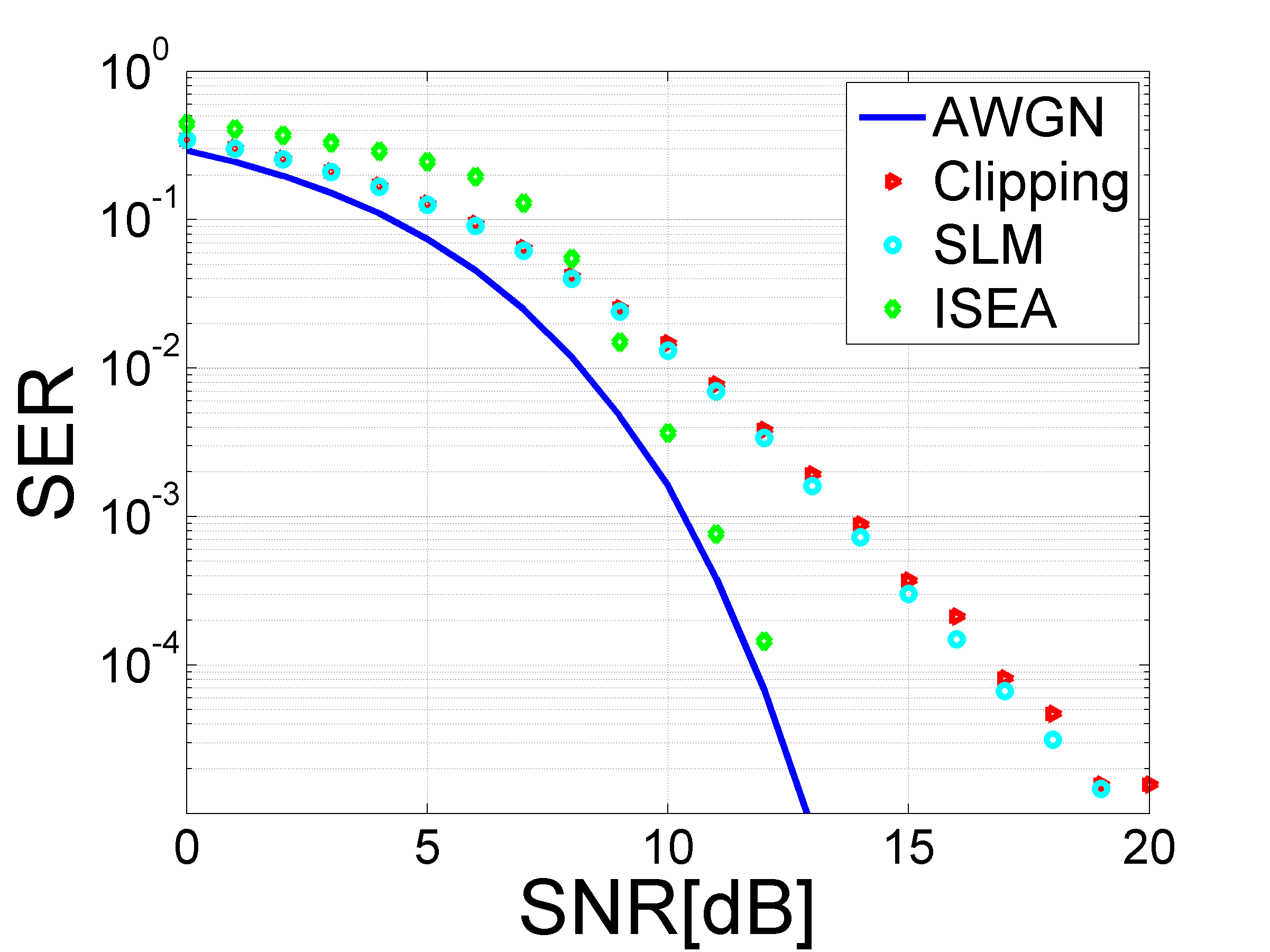}
            \caption{}
            \label{fig:SER_Vs_SNR_k_1_2_HQ QPSK}
        \end{subfigure}%
        ~
        \begin{subfigure}[b]{0.25\textwidth}
            \includegraphics[width=\textwidth]{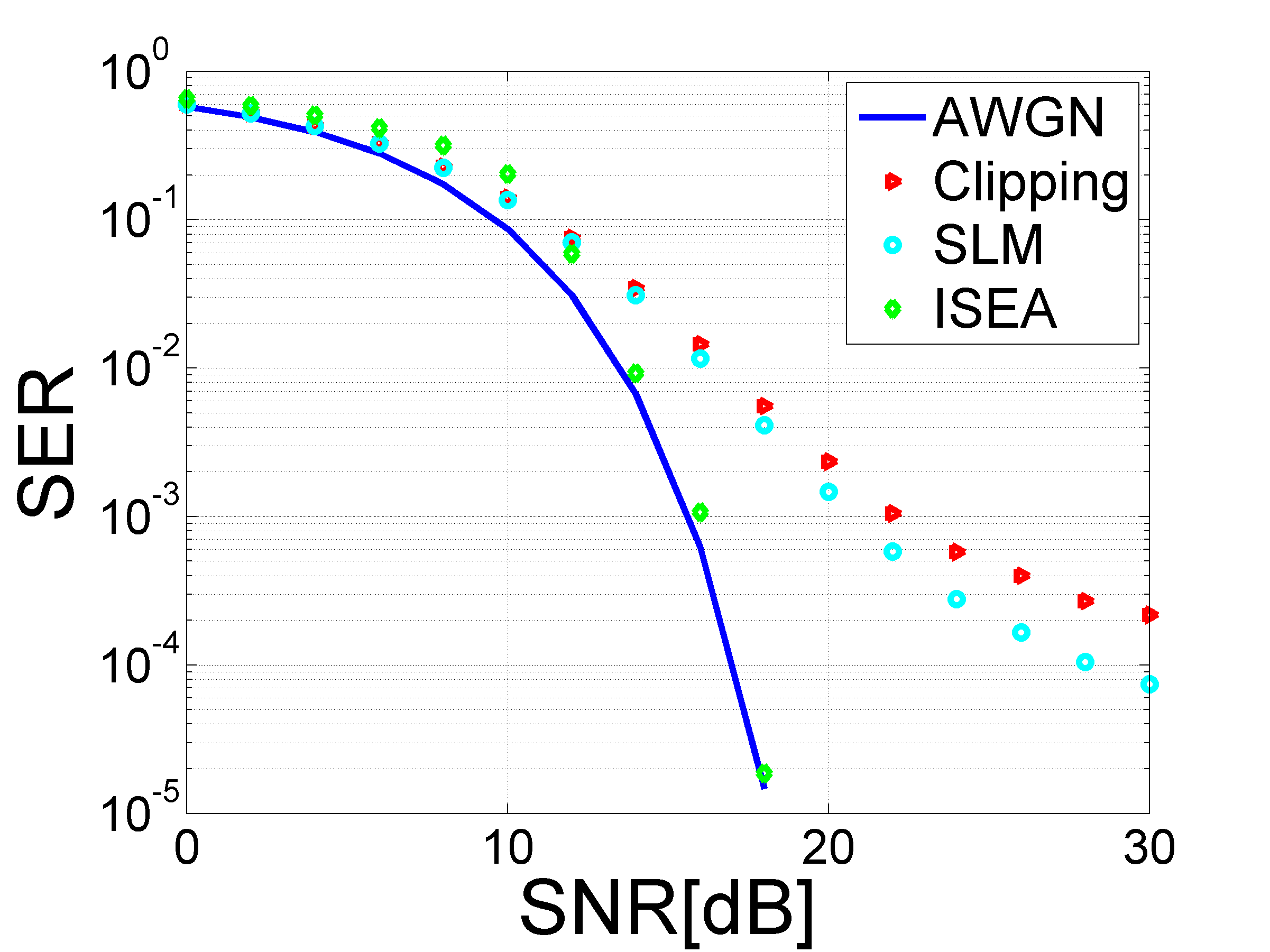}
            \caption{}
            \label{fig:SER_Vs_SNR_k_1_5_HQ 8PSK}
        \end{subfigure}
        \caption{Simulation results of SER vs. SNR with $N=1024$ and $\sigma_s^2=1$ based on averaging 2000 OFDM realizations, corresponding to a total of $\sim10^6$ symbols. (a) QPSK constellation with $\beta=3.35$[dB] ($\kappa=1.08$) (b) $8$-PSK constellation with $\beta=4.71$[dB] ($\kappa=1.4$) (c) QPSK constellation with $\beta=3.87$[dB] ($\kappa=1.2$) (d) $8$-PSK constellation with $\beta=5.12$[dB] ($\kappa=1.5$)}
        \label{fig:SER_Vs_SNR}
\end{figure}

We shall present results for two systems with an FFT size of $N=1024$, normalized signal power $\sigma_s^2=1$, and constellations of type QPSK and 8-PSK. 
First, we consider the case of a signal-independent-noiseless channel and examine clipping noise versus
AV noise. Figs. \ref{fig:constellationQPSK} and \ref{fig:constellation8PSK} show the constellation diagrams of the traditional and the proposed DCO-OFDM methods for QPSK and
$8$-PSK respectively. As seen, a relatively low DC bias with $\kappa=1$ and $\kappa=1.3$ is applied for QPSK and $8$-PSK respectively, and while the proposed method's estimators
(before the slicer) are exactly the true values, the clipped signal suffers from a significant distortion, which makes it much more vulnerable to signal-independent noise. We
remind that typical values for the DC bias used in traditional DCO-OFDM are at least two standard deviations ($\kappa\ge 2$) of the OFDM signal, and Figs.
\ref{fig:constellationQPSK} and \ref{fig:constellation8PSK} show the symbols estimators for QPSK with $\kappa=1$ and for $8$-PSK with $\kappa=1.3$, only $50\%$ and $65\%$
respectively of the minimal conventional value (which reduces the transmitted optical power by $60\%$ and $46\%$ respectively). Clearly, the proposed method achieves satisfactory
results with a significantly reduced DC bias. Fig. \ref{fig:Error_rate_SER0_SER_p_a} shows the SER, {defined as the number of incorrectly estimated symbols divided by the
total number of symbols}, as a function of the bias level with and without using the algorithm for QPSK and $8$-PSK. The numerically assessed value of $p_a$ is also shown in the
figure, together with the analytical curve obtained from Eq. \eqref{eq35}. The dotted vertical line indicates the bias level beyond which the ISEA algorithm converges to the exact
value (SER$=0$). The value of $p_{th}$ corresponds to the intersection of this dotted vertical line with the curve describing $p_a$. Notice that the values of $p_{th}$ are
$0.1355$ and $0.06654$ in the cases of QPSK and $8$-PSK, respectively. The corresponding respective values of $B_{DC}$ in Figs. \ref{fig:Error_rate_SER0_SER_p_a_QPSK} and
\ref{fig:Error_rate_SER0_SER_p_a_8PSK} are $1.1$ and $1.5$, and they can be checked to agree well with estimates based on Eq. \eqref{eq43}. Fig. \ref{fig:avg_num_of_iter} shows
the average number of iterations needed for the ISEA algorithm to converge as a function of the bias for QPSK and $8$-PSK. Here too, the vertical dotted lines indicate the
threshold bias level (as obtained from Figs. \ref{fig:Error_rate_SER0_SER_p_a_QPSK} and \ref{fig:Error_rate_SER0_SER_p_a_8PSK}). Evidently, when the algorithm converges to the
correct solution (i.e. over the threshold), it does so within only few iterations.

We now move on to assessing the performance of the ISEA algorithm in the presence of AWGN (i.e. with $\sigma_v^2>0$). In this case we compare the ISEA performance not only with the that of the standard clipping method, but also with the performance of the selective mapping (SLM) method, which combats clipping noise by means of PAPR reduction \cite{nadal2011comparison}, and was selected as a typical representative of the PAPR mitigation approach. The SLM method was implemented exactly as detailed in \cite{nadal2011comparison}, using $128$ alternative transmit sequences.\footnote{Each transmit sequence is generated via an element by element multiplication by a phase vector, whose $N$
elements are constrained to Hermitian symmetry, so that $N/2-1$ of them are chosen randomly (and uniformly) from the set $\{-1,+1\}$. Furthermore, we assume that the phase vector of the chosen lowest PAPR sequence is known to the receiver.}
This number is significantly higher than the typical numbers used in SLM implementations, and hence the SLM results that we report can be viewed as an effective upper-bound for this method's performance. The comparison of the ISEA approach with the SLM method will be relevant when considering the complexity implications in what follows. In addition, we also show the results for an AWGN channel that is not subject to the non-negativity constraint (where $y[n]=s[n]+v[n]$), and hence its SER constitutes a theoretical lower bound.

\begin{figure}[]
        \centering
        \begin{subfigure}[b]{0.25\textwidth}
            \includegraphics[width=\textwidth]{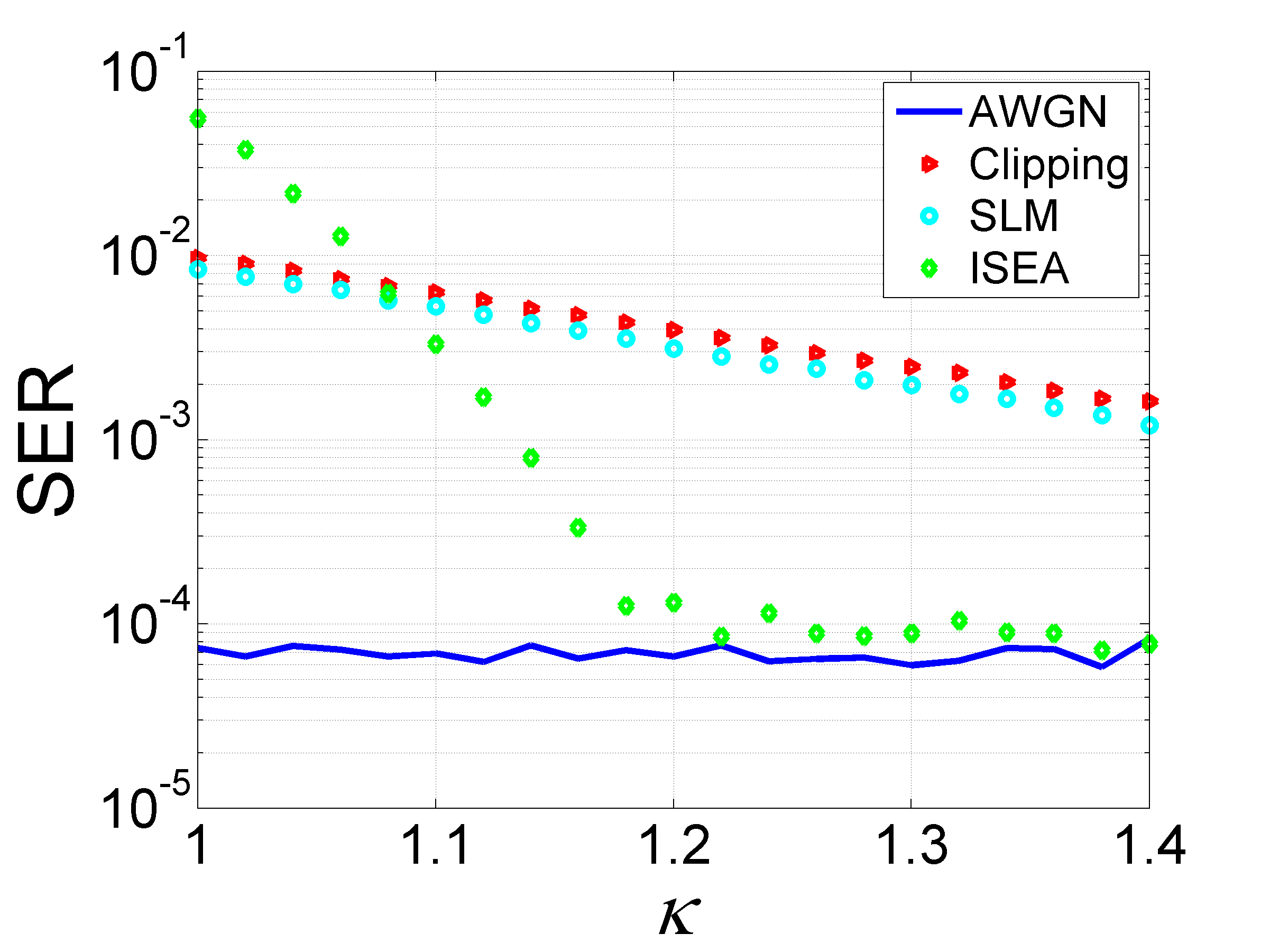}
            \caption{}
            \label{fig:SER_Vs_kappa_QPSK}
        \end{subfigure}%
        ~
        \begin{subfigure}[b]{0.25\textwidth}
           \includegraphics[width=\textwidth]{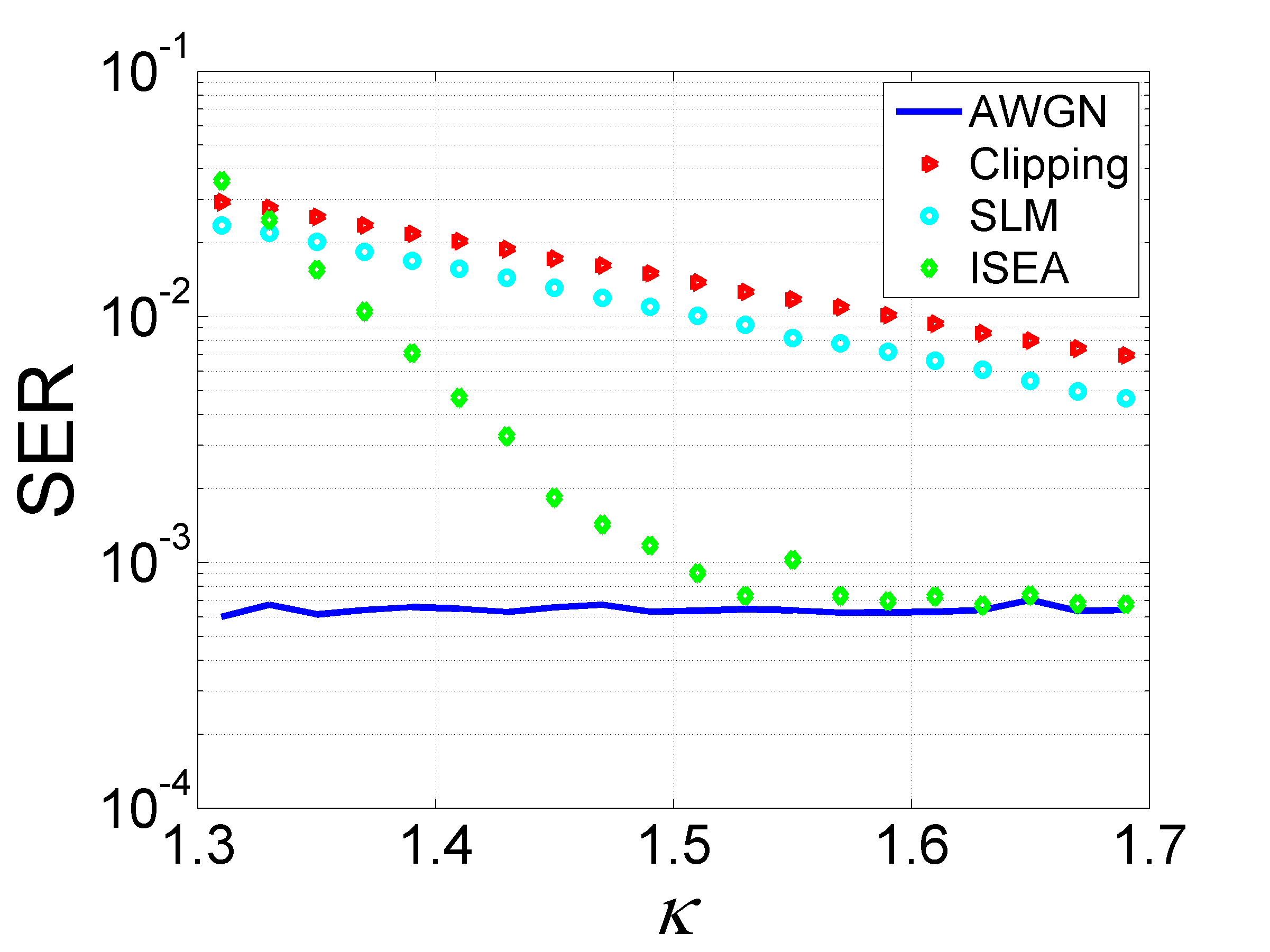}
            \caption{}
            \label{fig:SER_Vs_kappa_8PSK}
        \end{subfigure}
        \caption{Simulation results of SER vs. $\kappa$ for fixed SNR with $N=1024$ and $\sigma_s^2=1$ based on averaging 2000 OFDM realizations, corresponding to a total of $\sim10^6$ symbols. (a) QPSK constellation with SNR$=12$[dB] (b) $8$-PSK constellation with SNR$=16$[dB]}
        \label{fig:SER_Vs_kappa}
\end{figure}

\begin{figure}[t]
\centering
    \includegraphics[width=0.3\textwidth]{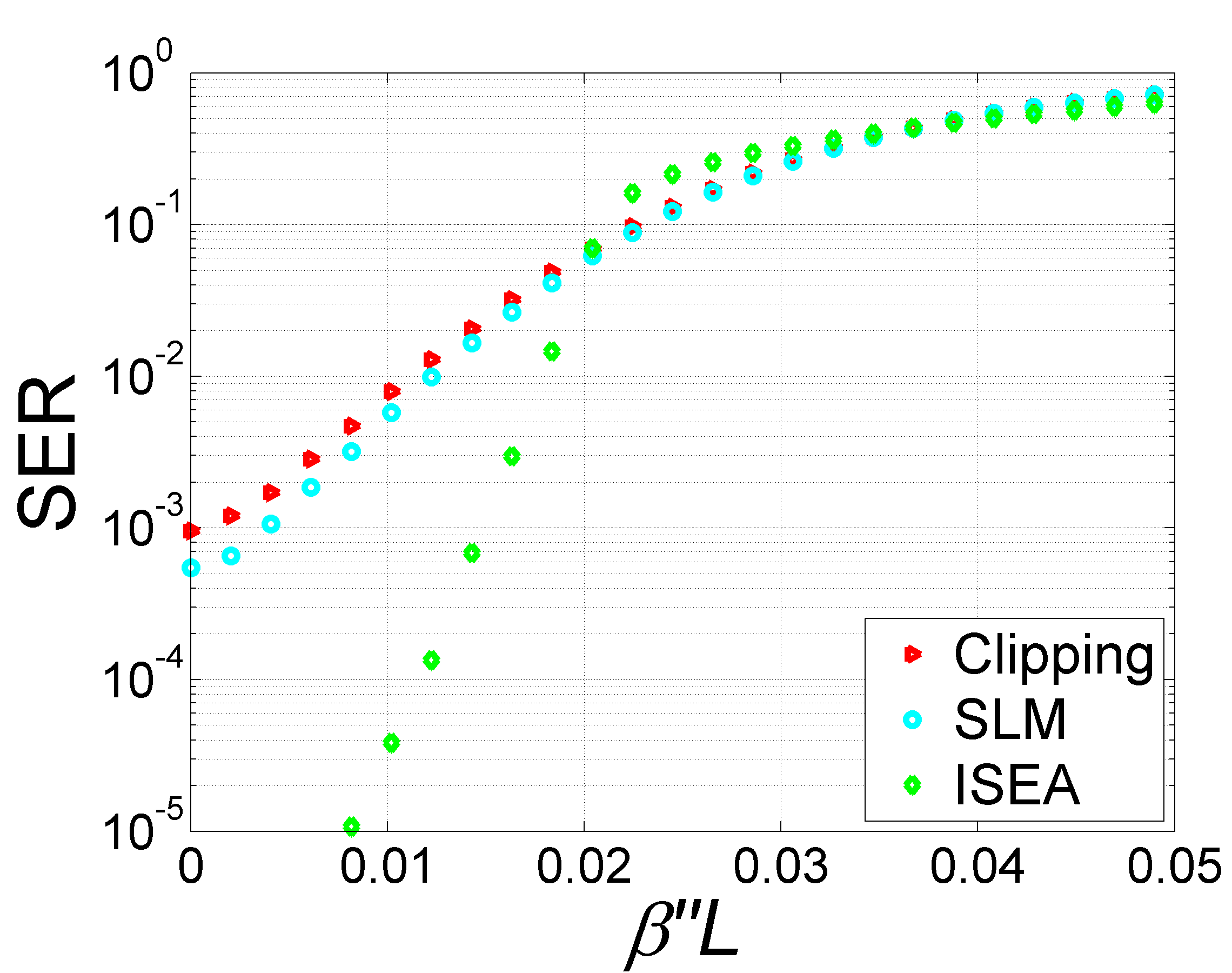}
    \caption{{Simulation results of SER vs. the normalized dispersion $B^2\beta'' L$ where $B$ is the channel bandwidth, $\beta''$ is the dispersion coefficient, and $L$ is the length of the link. The curves were obtained with $N=1024$, $\sigma_s^2=1$, $\kappa=1.6$ and SNR$=20$[dB] in $8$-PSK constellation. Results are based on averaging 2000 OFDM realizations, corresponding to a total of $\sim10^6$ symbols.}}
    \label{fig:SER_vs_disperion_constant_8PSK}
\end{figure}

We now examine the dependence of the SER on the SNR in the cases of QPSK and $8$-PSK transmission, with the bias levels set slightly below and slightly above the corresponding
thresholds (recall that the threshold $\kappa$ is $1.1$ in the case of QPSK and $1.4$ with $8$-PSK, as seen in Fig. \ref{fig:avg_num_of_iter}). Figs.
\ref{fig:SER_Vs_SNR_k_1_08_HQ_QPSK} and \ref{fig:SER_Vs_SNR_k_1_4_HQ_8PSK} show the SER versus the SNR [defined as $10\log_{10}(\sigma_s^2/\sigma_v^2)$] slightly below the
$\kappa$-threshold, whereas in Figs. \ref{fig:SER_Vs_SNR_k_1_2_HQ QPSK}--\ref{fig:SER_Vs_SNR_k_1_5_HQ 8PSK} $\kappa$ is slightly higher than the threshold value. When $\kappa$ is
set slightly below the threshold (Figs. \ref{fig:SER_Vs_SNR_k_1_08_HQ_QPSK} and \ref{fig:SER_Vs_SNR_k_1_4_HQ_8PSK}), the performance of the ISEA and {the two other}
clipping{-based} methods are comparable in the low SNR regime, but in the relevant range, with the SER set to $10^{-3}$ the ISEA scheme gains relative to the other methods. With
QPSK the ISEA result is similar to that of SLM and it is $\sim1$[dB] better in SNR than in the case of standard clipping. With 8-PSK the SNR gain with respect to standard clipping
and SLM, is  $8$[dB] and $4$[dB], respectively. When $\kappa$ is slightly above the threshold value (Figs. \ref{fig:SER_Vs_SNR_k_1_2_HQ QPSK} and \ref{fig:SER_Vs_SNR_k_1_5_HQ
8PSK}), the performance of the ISEA method rapidly converges to the performance of the unconstrained AWGN channel, implying that the positivity requirement does not introduce any
penalty. Figure \ref{fig:SER_Vs_kappa} shows the SER versus $\kappa$ for a fixed SNR value, illustrating the difference between the various methods below and above the threshold
value of $\kappa$. Note that unlike in the case of the other two methods, the performance of the ISEA scheme rapidly approaches the theoretical lower bound. Table
\ref{table:kappaSNRfixedSER} shows the threshold values of $\kappa$ and the gain in SNR with respect to standard bias and clipping with the SER set to $10^{-3}$ for a few common
constellations. As seen, the proposed method outperforms clipping in SNR requirement while reducing the DC bias. It is noteworthy that the DC bias threshold was computed for a
noiseless channel, hence the threshold in a noisy channel would most likely be higher for low SNRs and would approach the noiseless threshold as the SNR increases. This explains
the degraded performance of the proposed method for low SNRs, expressed by the gap between the lower bound and ISEA's curves in the low SNR region of Fig.
\ref{fig:SER_Vs_SNR_k_1_2_HQ QPSK} and \ref{fig:SER_Vs_SNR_k_1_5_HQ 8PSK}.


\begin{table}[t]
\centering
    \begin{tabular}{ | c | c | c |}
    \hline
    Constellation & $\kappa$ & SNR Gain [dB] \\ \hline
    QPSK & $1.2$ & $3$ \\ \hline
    $8$-PSK & $1.5$ & $6.1$ \\ \hline
    $16$-PSK & $2.1$ & $4.2$ \\ \hline
    $16$-QAM & $2.1$ & $1.47$ \\ \hline
    \end{tabular}
    \caption{Threshold values of $\kappa$ and the SNR gain, with respect to the standard DCO-OFDM method, for a fixed SER of $10^{-3}$, with $N=1024$ and $\sigma_s^2=1$. In all cases the ISEA method outperforms clipping.}
\label{table:kappaSNRfixedSER}
\end{table}

\begin{figure}[t]
\centering
    \includegraphics[width=0.5\textwidth]{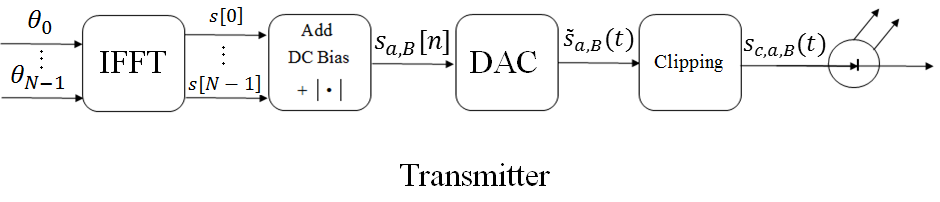}
    \caption{A block diagram of the simplified transmitter, where DC bias addition and the absolute value operator are applied on the digital signal, such that $s_{a,B}[n]=\left|s[n]+B_{DC}\right|$.}
    \label{fig:Simplified_Block_Diagram}
\end{figure}


In Fig. \ref{fig:SER_vs_disperion_constant_8PSK} we compare the tolerance of the three methods to chromatic dispersion. It should be noted in this context that schemes based on direct laser modulation, such as the ones that we are considering, are not intended for highly dispersive links. Yet, comparing the three schemes with respect to this phenomenon can be viewed as a plausible indicator of their relative tolerance to inter-symbol-interference mechanisms in a more general sense. Since the absolute value operation underpinning the ISEA approach produces larger spectral broadening than standard-clipping or SLM, the ISEA method is expected to be less tolerant to dispersion. Indeed, this reality is seen in Fig. \ref{fig:SER_vs_disperion_constant_8PSK}, where the rate at which the ISEA performance deteriorates with increasing dispersion is higher than it is with the other two approaches. Nonetheless, the vast advantage of the ISEA approach more than compensates for this sensitivity.

{Finally, we wish to address the implications of the proposed ISEA method with respect to the complexity of the overall system. The receiver of the ISEA scheme involves one FFT
and one IFFT operation per iteration and hence the complexity increase in comparison with the standard clipping-based DCO-OFDM approach, is determined by the number of iterations
that are required for the algorithm to converge. As witnessed by the simulation results displayed in Fig. \ref{fig:avg_num_of_iter}, the average of the number of iterations needed
for convergence (above threshold) is practically equal to $3$ and its standard deviation is between one order of magnitude and two orders of magnitude smaller.\footnote{This was
also confirmed in simulations with a broad variety of system set-ups.} This implies that the effective complexity of the ISEA method is larger by a factor of 6 relative to
standard clipping, where the receiver performs a single FFT operation. It is interesting to compare this cost, with the cost of a typical PAPR reduction method, such as SLM
\cite{nadal2011comparison}. The SLM method also involves additional IFFT operations, whose number equals the number of alternative transmit sequences that it includes. Hence, an
SLM method with the same complexity overhead as the ISEA would include 6 alternative transmit sequences, and its performance would be inferior to that of the SLM results that we
have shown in Figs. \ref{fig:SER_Vs_SNR}--\ref{fig:SER_vs_disperion_constant_8PSK}, where the number of alternative transmit sequences was 128. Although we have only considered
SLM explicitly, it must be noted that SLM is not considerably different from other PAPR reduction methods in terms of its complexity performance tradeoff.\footnote{Other relevant
methods in this context are; partial transmit sequence, interleaving, tone reservation, and tone injection. As detailed in \cite{nadal2011comparison},\cite{jiang2008overview}
these methods also involve additional FFT/IFFT operations and/or solvers of optimization problems.} All these methods aim at reducing the PAPR, as opposed to the ISEA method which
\textit{copes} with it. In this sense, the PAPR reduction methods are complimentary to the ISEA approach and in principle the two can be combined with one another.

We note in passing, that assuming synchronization, a scheme with the same receiver but with a different, significantly easier to implement transmitter will yield exactly the same
results presented above. The simplified transmitter, proposed in Fig. \ref{fig:Simplified_Block_Diagram}, differs from the original by applying the DC bias and the absolute value
operation digitally on the samples, followed by simple clipping after the DAC, to ensure non-negativity. In this manner, all the values at the samples instances are identical to
those produced by the transmitter presented in Fig. \ref{fig:Block_Diagram}, hence leading to identical results.

\section{Conclusions}
We proposed a new approach for meeting the non-negativity constraint in IM/DD. The new scheme, combines the absolute value operator and the ISEA algorithm, and shows a significant
improvement over the standard DCO-OFDM method, both in terms of SER performance for a given DC bias and in terms of the bias needed for a given required SER. With a relatively
moderate DC bias-index of $3.87$[dB] and $5.12$[dB] for QPSK and $8$-PSK respectively, the theoretical lower bound of a standard AWGN channel is achieved. Along with a rudimentary
analysis of the AV noise, both analytically approximated lower bound on the DC bias and simulation results of the proposed method have been presented. {Our analysis was
performed for the case of a linear system, neglecting component imperfections and hence provides an upper bound for the system performance. The consideration of the tolerance of
the scheme for system non-idealities is left for future study.}

\bibliographystyle{unsrt}
\bibliography{Bibfile}

\begin{thebibliography}{10}
\providecommand{\url}[1]{#1}
\csname url@samestyle\endcsname
\providecommand{\newblock}{\relax}
\providecommand{\bibinfo}[2]{#2}
\providecommand{\BIBentrySTDinterwordspacing}{\spaceskip=0pt\relax}
\providecommand{\BIBentryALTinterwordstretchfactor}{4}
\providecommand{\BIBentryALTinterwordspacing}{\spaceskip=\fontdimen2\font plus
\BIBentryALTinterwordstretchfactor\fontdimen3\font minus
  \fontdimen4\font\relax}
\providecommand{\BIBforeignlanguage}[2]{{%
\expandafter\ifx\csname l@#1\endcsname\relax
\typeout{** WARNING: IEEEtran.bst: No hyphenation pattern has been}%
\typeout{** loaded for the language `#1'. Using the pattern for}%
\typeout{** the default language instead.}%
\else
\language=\csname l@#1\endcsname
\fi
#2}}
\providecommand{\BIBdecl}{\relax}
\BIBdecl

\bibitem{chang1970orthogonal}
R.~W. Chang, ``Orthogonal frequency multplex data transmission system,'' Jan.~6
  1970, uS Patent 3,488,445.

\bibitem{salz1969fourier}
J.~Salz and S.~Weinstein, ``Fourier transform communication system,'' in
  \emph{Proceedings of the first ACM symposium on Problems in the optimization
  of data communications systems}.\hskip 1em plus 0.5em minus 0.4em\relax ACM,
  1969, pp. 99--128.

\bibitem{peled1980frequency}
A.~Peled and A.~Ruiz, ``Frequency domain data transmission using reduced
  computational complexity algorithms,'' in \emph{Acoustics, Speech, and Signal
  Processing, IEEE International Conference on ICASSP'80.}, vol.~5.\hskip 1em
  plus 0.5em minus 0.4em\relax IEEE, 1980, pp. 964--967.

\bibitem{dixon2001orthogonal}
B.~J. Dixon, R.~D. Pollard, and S.~Iezekiel, ``Orthogonal frequency-division
  multiplexing in wireless communication systems with multimode fiber feeds,''
  \emph{Microwave Theory and Techniques, IEEE Transactions on}, vol.~49, no.~8,
  pp. 1404--1409, 2001.

\bibitem{armstrong2009ofdm}
J.~Armstrong, ``Ofdm for optical communications,'' \emph{Journal of lightwave
  technology}, vol.~27, no.~3, pp. 189--204, 2009.

\bibitem{armstrong2006power}
J.~Armstrong and A.~Lowery, ``Power efficient optical ofdm,'' \emph{Electronics
  Letters}, vol.~42, no.~6, pp. 370--372, 2006.

\bibitem{armstrong2008comparison}
J.~Armstrong and B.~Schmidt, ``Comparison of asymmetrically clipped optical
  ofdm and dc-biased optical ofdm in awgn,'' \emph{Communications Letters,
  IEEE}, vol.~12, no.~5, pp. 343--345, 2008.

\bibitem{carruthers1996multiple}
J.~B. Carruthers and J.~M. Kahn, ``Multiple-subcarrier modulation for
  nondirected wireless infrared communication,'' \emph{Selected Areas in
  Communications, IEEE Journal on}, vol.~14, no.~3, pp. 538--546, 1996.

\bibitem{gonzalez2005ofdm}
O.~Gonzalez, R.~Perez-Jimenez, S.~Rodriguez, J.~Rabad{\'a}n, and A.~Ayala,
  ``Ofdm over indoor wireless optical channel,'' in \emph{Optoelectronics, IEE
  Proceedings-}, vol. 152, no.~4.\hskip 1em plus 0.5em minus 0.4em\relax IET,
  2005, pp. 199--204.

\bibitem{li2007channel}
X.~Li, R.~Mardling, and J.~Armstrong, ``Channel capacity of im/dd optical
  communication systems and of aco-ofdm,'' in \emph{Communications, 2007.
  ICC'07. IEEE International Conference on}.\hskip 1em plus 0.5em minus
  0.4em\relax IEEE, 2007, pp. 2128--2133.

\bibitem{tsonev2012novel}
D.~Tsonev, S.~Sinanovic, and H.~Haas, ``Novel unipolar orthogonal frequency
  division multiplexing (u-ofdm) for optical wireless,'' in \emph{Vehicular
  Technology Conference (VTC Spring), 2012 IEEE 75th}.\hskip 1em plus 0.5em
  minus 0.4em\relax IEEE, 2012, pp. 1--5.

\bibitem{barrami2014optical}
F.~Barrami, Y.~Le~Guennec, E.~Novakov, and P.~Busson, ``An optical power
  efficient asymmetrically companded dco-ofdm for im/dd systems,'' in
  \emph{Wireless and Optical Communication Conference (WOCC), 2014 23rd}.\hskip
  1em plus 0.5em minus 0.4em\relax IEEE, 2014, pp. 1--6.

\bibitem{svaluto2010novel}
M.~Svaluto~Moreolo, R.~Mu{\~n}oz, and G.~Junyent, ``Novel power efficient
  optical ofdm based on hartley transform for intensity-modulated
  direct-detection systems,'' \emph{Journal of lightwave Technology}, vol.~28,
  no.~5, pp. 798--805, 2010.

\bibitem{tang2003synchronization}
H.~Tang, K.~Y. Lau, and R.~W. Brodersen, ``Synchronization schemes for packet
  ofdm system,'' in \emph{Communications, 2003. ICC'03. IEEE International
  Conference on}, vol.~5.\hskip 1em plus 0.5em minus 0.4em\relax IEEE, 2003,
  pp. 3346--3350.

\bibitem{jin2011optical}
X.~Jin and J.~Tang, ``Optical ofdm synchronization with symbol timing offset
  and sampling clock offset compensation in real-time imdd systems,''
  \emph{Photonics Journal, IEEE}, vol.~3, no.~2, pp. 187--196, 2011.

\bibitem{armstrong2006spc07}
J.~Armstrong, B.~Schmidt, D.~Kalra, H.~A. Suraweera, and A.~J. Lowery,
  ``Spc07-4: Performance of asymmetrically clipped optical ofdm in awgn for an
  intensity modulated direct detection system,'' in \emph{Global
  Telecommunications Conference, 2006. GLOBECOM'06. IEEE}.\hskip 1em plus 0.5em
  minus 0.4em\relax IEEE, 2006, pp. 1--5.

\bibitem{svaluto2010power}
M.~Svaluto~Moreolo, ``Power efficient and cost-effective solutions for optical
  ofdm systems using direct detection,'' in \emph{Transparent Optical Networks
  (ICTON), 2010 12th International Conference on}.\hskip 1em plus 0.5em minus
  0.4em\relax IEEE, 2010, pp. 1--4.

\bibitem{kim1999clipping}
D.~Kim and G.~L. Stuber, ``Clipping noise mitigation for ofdm by decision-aided
  reconstruction,'' \emph{Communications Letters, IEEE}, vol.~3, no.~1, pp.
  4--6, 1999.

\bibitem{nadal2011comparison}
L.~Nadal, M.~S. Moreolo, J.~M. Fabrega, and G.~Junyent, ``Comparison of peak
  power reduction techniques in optical ofdm systems based on fft and fht,'' in
  \emph{Transparent Optical Networks (ICTON), 2011 13th International
  Conference on}.\hskip 1em plus 0.5em minus 0.4em\relax IEEE, 2011, pp. 1--4.

\bibitem{jiang2008overview}
T.~Jiang and Y.~Wu, ``An overview: peak-to-average power ratio reduction
  techniques for ofdm signals,'' \emph{IEEE transactions on broadcasting},
  vol.~54, no.~2, p. 257, 2008.

\end{thebibliography}

\end{document}